\documentclass[a4paper,11pt]{article}
\usepackage{jheppub}
\usepackage{amsmath}
\usepackage{amsthm}
\usepackage{tabularx,graphicx}
\usepackage{epstopdf}
\usepackage{makecell}
\usepackage{latexsym}
\usepackage{amssymb}
\usepackage[dvipsnames]{xcolor}
\usepackage{colortbl}
\usepackage{bbm}
\usepackage{bm}
\usepackage{titlesec}
\usepackage{dsfont}
\usepackage{feynmp}
\usepackage{slashed}
\usepackage{multirow}
\usepackage{placeins}
\usepackage{url}

\newcommand{\mc}[1]{\mathcal{#1}}

\newcommand{\beq}{\begin{eqnarray}}
	\newcommand{\eeq}{\end{eqnarray}}

\newcommand{\bsp}{\begin{aligned}}
	\newcommand{\esp}{\end{aligned}}

\newcommand{\hc}{{\rm h.c.}}

\definecolor{darkblue}{rgb}{0.,0.,0.4}
\definecolor{darkred}{rgb}{0.5,0.,0.}
\definecolor{BlueViolet}{RGB}{138,43,226}
\definecolor{SkyBlue}{RGB}{30,144,255}
\definecolor{DarkGreen}{RGB}{0,100,0}

\usepackage{mathtools}

\usepackage{tikz-cd}
\usepackage{booktabs}

\usepackage[normalem]{ulem}

\usepackage{mathrsfs}
\usepackage{comment}

\newcommand{\Z}{\mathbb{Z}}

\usepackage{svg}

\usetikzlibrary{calc}

\usepackage[mathscr]{eucal}
\makeatletter
\def\instring#1#2{TT\fi\begingroup
  \edef\x{\endgroup\noexpand\in@{#1}{#2}}\x\ifin@}
\def\isuppercase#1{%
  \instring{#1}{ABCDEFGHIJKLMNOPQRSTUVWXYZ}%
}%
\newcommand{\C@lIfUpper}[1]{
 \if\isuppercase{#1}\mathscr{#1}%
 \else #1%
 \fi
}
\newcommand{\cat}[1]{\mathit{\@tfor\next:=#1\do{\C@lIfUpper{\next}}}}
\makeatother

\usepackage{xparse}

\newcommand{\U}{\mathrm U}
\newcommand{\Spin}{\mathrm{Spin}}

\newcommand{\SU}{\mathrm{SU}}

\newcommand{\SO}{\mathrm{SO}}

\newcommand{\bl}{\text{--}}

\newcommand{\SH}{\mathrm{SH}}
\newcommand{\rH}{\mathrm{H}}

\newcommand{\tVect}{\mathbf{2Vect}}
\newcommand{\tsVect}{\mathbf{2sVect}}

\newcommand{\Ahat}{\widehat{A}}

\newcommand{\term}{\emph}

\newcommand{\fB}{\mathfrak B}

\newcommand{\alignedintertext}[1]{%
  \noalign{%
    \vskip\belowdisplayshortskip
    \vtop{\hsize=0.75\linewidth#1\par
    \expandafter}%
    \expandafter\prevdepth\the\prevdepth
  }%
}

\newcommand\MAILTO[1]{\href{mailto:#1}{\nolinkurl{#1}}}

\DeclareDocumentCommand{\shortexact}{s O{} O{} mmmm}{
\IfBooleanTF{#1}{ 
\begin{tikzcd}[ampersand replacement=\&]
	{1} \& {#4} \& {#5} \& {#6} \& {1#7}
	\arrow[from=1-1, to=1-2]
	\arrow["#2", from=1-2, to=1-3]
	\arrow["#3", from=1-3, to=1-4]
	\arrow[from=1-4, to=1-5]
\end{tikzcd}
}{ 
\begin{tikzcd}[ampersand replacement=\&]
	{0} \& {#4} \& {#5} \& {#6} \& {0#7}
	\arrow[from=1-1, to=1-2]
	\arrow["#2", from=1-2, to=1-3]
	\arrow["#3", from=1-3, to=1-4]
	\arrow[from=1-4, to=1-5]
\end{tikzcd}
}}

\usepackage{xspace}

\makeatletter
\renewcommand{\p@subsection}{}
\makeatother

\usepackage[capitalize, noabbrev]{cleveref}

\newtheorem{thm}{Theorem}

\newtheorem{lem}{Lemma}

\newtheorem*{thm*}{Theorem}

\theoremstyle{definition}

\newtheorem{example}{Example}
\theoremstyle{remark}

\crefname{cor}{Corollary}{Corollaries}

\crefname{thm}{Theorem}{Theorems}
\Crefname{thm}{Theorem}{Theorems}
\crefname{prop}{Proposition}{Propositions}
\crefname{lem}{Lemma}{Lemmas}
\crefname{defn}{Definition}{Definitions}

\title{Symmetric Gapped States and Symmetry-Enforced Gaplessness in 3 Dimensions}

\author[a]{Arun Debray,}
\affiliation[a]{Department of Mathematics, University of Kentucky, 719 Patterson Office Tower, Lexington, KY 40506-0027}
\emailAdd{a.debray@uky.edu}

\author[b]{Matthew Yu,}
\affiliation[b]{School of Natural Sciences, Institute for Advanced Study, Princeton, NJ, USA}
\emailAdd{yumatthew70@gmail.com}

\author[c]{and Weicheng Ye}
\affiliation[c]{Department of Physics and Astronomy, and Stewart Blusson Quantum Matter Institute, University of British Columbia, Vancouver, BC, Canada V6T 1Z1}
\emailAdd{victoryeofphysics@gmail.com}

\abstract{
We establish a comprehensive framework for constraining the symmetry-preserving infrared (IR) possibilities of a fermionic quantum theory in three spatial dimensions using its \textit{quantum anomalies} associated with a finite symmetry.
We uncover a fundamental dichotomy among these anomalies: the first class of anomalies can \textit{always} be realized by symmetric gapped states, while the second class can \textit{never} be realized by gapped states without breaking the given symmetry, establishing the phenomenon of symmetry-enforced gaplessness in these settings. Moreover, using the construction of symmetry extension, we construct the candidate gapped states that theories with the first class of anomalies can flow to in the IR. As an application, we provide concrete predictions of the candidate IR phases of (3+1)-dimensional gauge theories based on our results. Our results also suggest that systems with certain discrete chiral anomalies cannot be gapped out by adding arbitrary bosonic degrees of freedom.
}

\notoc
\compress
\begin{document}

\maketitle
\newpage

\section{Introduction}\label{sec:introduction}
Symmetries\footnote{For this paper, all symmetries are 0-form on-site symmetries. Moreover, we will restrict ourselves to finite symmetries unless explicitly stated otherwise.} play a central role in modern physics \cite{wen2004quantum}, most notably within the Landau paradigm of phases and phase transitions. Nevertheless,
it is recognized that symmetries alone are insufficient to characterize all possible quantum phases. Thus, considerable effort has been devoted to expanding the landscape beyond the traditional framework \cite{2019PhR...827....1S,Wang:2022ucy,Bhardwaj:2023kri,Schafer-Nameki:2023jdn,Bhardwaj:2023fca,Bhardwaj:2024qiv,Bhardwaj:2025piv,Luo:2023ive,Stockall:2025ngz}. In this broader perspective, \textit{quantum anomalies}\footnote{In this work, we restrict attention to anomalies other than gauge anomalies, which indicate a fundamental incompatibility with gauge invariance at the quantum level. In contrast, quantum anomalies considered here are always associated with \textit{global symmetries}. These quantum anomalies should be thought of as intrinsic properties of the theory and do not suggest that the theory is pathological. Moreover, while quantum anomalies can naturally be canceled by realizing the theory on the boundary of a symmetry-protected topological (SPT) state, the system under study does not necessarily require such a boundary realization, as long as the symmetries responsible for the anomalies are either emergent or non-on-site.}, also referred to as \textit{'t Hooft anomalies} \cite{Hooft1980}, have emerged as a powerful diagnostic tool.

\textit{Quantum anomalies} are commonly defined as obstructions to gauging a global symmetry \cite{Hooft1980,peskin2018introduction}. As such, their presence immediately implies that the corresponding quantum state cannot be completely trivial. 
Assuming that the symmetry remains unbroken in the infrared (IR), this basic constraint suggests that either the IR theory is gapless, or it realizes a nontrivial topological order. 

It is natural to ask whether anomalies alone can provide further insight into the physical system. In particular, can anomalies be strong enough to rule out \emph{all} symmetric gapped states, including those with intrinsic topological order?\footnote{We will use the terms \emph{gapped states} and \emph{topological orders} interchangeably, and exclude more exotic possibilities such as fracton states. We assume that a topological order is described by a unitary TQFT $Z$ with $\dim Z(S^3)=1$. Moreover, these gapped states all take the form of finite gauge theories in 3 spatial dimensions \cite{Lan_2019,JF,Decoppet:2025eic}.} This phenomenon is termed \textit{symmetry-enforced gaplessness}\footnote{Sometimes it is also called \textit{anomaly-enforced gaplessness}, e.g., in Ref.~\cite{Brennan:2023kpo}.} in Ref.~\cite{2014PhRvB..89s5124W}, and has proven to be crucial in the understanding of deconfined quantum critical points \cite{Wang:2017txt} and has been extensively analyzed, e.g., in Refs.~\cite{CO1,CO2,Brennan:2023kpo,Brennan:2023ynm,Yang:2023gvi,Cheng:2024awi,Ning:2021kqc,Ye:2022bkx,Debray:2023bosonization,ZJSNew}. Conversely, given a prescribed anomaly, can one systematically construct candidate gapped states that realize the given anomaly? These candidate states may naturally serve as the candidate IR phases for a (3+1)-dimensional gauge theory or a 3-dimensional lattice system. Thus, answering these questions may give exact nonperturbative constraints on these strongly coupled systems.

For bosonic systems, an affirmative answer to the latter question is provided by symmetry extension constructions developed in Refs.~\cite{2014arXiv1404.3230K,2014PhRvL.112w1602K,Thorngren:2015gtw,Wang:2017loc,Tachikawa:2017gyf}, when the quantum anomaly takes values in ordinary cohomology \cite{Tachikawa:2017gyf}. In that framework, one first enlarges the symmetry group such that the anomaly becomes trivial upon extending to the larger group, and subsequently gauges part of the extended symmetry to obtain a gauge theory that faithfully realizes the original anomalous symmetry. This construction gives a remarkably general method for engineering symmetric gapped states that realize a given anomaly in bosonic systems.

In this work, we aim to generalize this framework to fermionic systems with finite symmetries. While attempts have been made to extend lattice \cite{Kobayashi:2019lep,Loo:2025,2021PhRvR...3a3056C} and path-integral constructions to fermionic systems, both approaches present significant technical challenges. Nevertheless, at our disposal is a category-theory-based framework that has recently been developed in \cite{Johnson-Freyd:2021tbq,D11,D8,Decoppet:2023bay,Decoppet:2024htz,Decoppet:2025eic,Debray:2025kfg}. In the companion paper \cite{Debray:2025kfg}, we provide a systematic construction of fermionic topological orders for a given anomaly. The formal procedure is similar to the bosonic case, but one notable exception is that one replaces ordinary cohomology with \textit{supercohomology} \cite{Wang:2017moj}. Moreover, we identify the precise condition under which such constructions are obstructed. This leads to a sharp refinement of the predictions from an anomaly in the form of symmetry-enforced gaplessness.

Our goal in this paper is to test this framework on interacting 
(3+1)d systems, especially systems with chiral fermions, which frequently have nonzero 't Hooft anomalies for finite global symmetries. We specifically consider a theory with a single Dirac fermion transforming in a representation $R$ of a gauge group $\mathcal G$, for which the ABJ anomaly breaks the $\U(1)^F$ chiral symmetry to a finite cyclic group~\cite[Eq.~(8)]{CO2}. We offer the following examples as tests of our framework.
\begin{enumerate}
    \item\label{case1} For $\mathcal G = \SU(2)$ and $R$ the adjoint representation, $\U(1)^F$ breaks to $\Z_8^F$. The symmetries, anomalies, and low-energy behavior of adjoint $\mathrm{QCD}_4$ have been extensively studied~\cite[Sec.~2.2]{Cordova:2018acb}. Notably, Córdova--Ohmori~\cite{CO2} showed that the IR theory of $\SU(2)$ adjoint QCD either breaks the $\Z_8^F$ symmetry or is gapless. See also Wan--Wang~\cite{Wan:2018djl}. We therefore include this example to illustrate how the results of~\cite{CO2,Cordova:2018acb,Wan:2018djl} are special cases of our more general approach.
    
    
    Specifically, we show that the anomaly of $\SU(2)$ adjoint QCD is not an in-supercohomology SPT; hence there is no (3+1)d fermionic topological order with the same anomaly. Therefore, either the $\Z_8^F$ symmetry breaks at low energy or the low-energy limit is gapless, in agreement with~\cite{CO2,Cordova:2018acb,Wan:2018djl}.
    \item\label{case2} For $\mathcal G = \SU(2N+1)$ and $R$ the fundamental representation, $\U(1)^F$ breaks to $\Z_2^F$. The anomaly is an element of the group of $(4+1)$-dimensional $\Z_2^F$-symmetric SPTs, which vanishes. Thus this $\Z_2^F$ symmetry is anomaly-free, so a trivial symmetric gapped state is not excluded by this anomaly.
    \item\label{case3} For $\mathcal G = \SO(2N+1)$ and $R$ the vector representation, $\U(1)^F$ breaks to $\Z_4^F$, which in (3+1)d has a $\Z_{16}$-valued anomaly. We find that, for all $N$, the anomaly of this theory is an in-supercohomology SPT, and in fact is realized as the anomaly of a $\Z_4$ gauge theory.\footnote{For $N = 1$, this theory can be obtained from $\SU(2)$ adjoint $\mathrm{QCD}_4$ by gauging the electric one-form $\Z_2$ symmetry~\cite[\S 4.2]{GKSW}. So this example and the example in case~\ref{case1} are related.}
    \item\label{case4} For $\mathcal G = \mathrm F_4$ and $R$ the fundamental representation, $\U(1)^F$ breaks to $\Z_{12}^F$, and the anomaly is a beyond-supercohomology SPT, so that in the IR, either $\Z_{12}^F$ breaks or the theory is gapless.
    \item\label{case5} We also consider a single left-handed Weyl fermion coupled to a bosonic field, each carrying charge $1$ under a $\Z_4$ symmetry. The full fermionic symmetry group is $\Z_2^F\times\Z_4$, and we find that this theory has a beyond-supercohomology anomaly. Thus, once again, we conclude that in the IR, either the $\Z_2^F\times\Z_4$ symmetry is broken, or the theory is gapless.
\end{enumerate}
In cases~\ref{case1}--\ref{case4}, the symmetry is $\Z_{2m}^F$ for some $m$. For these symmetry groups, Córdova--Ohmori~\cite{CO2} showed that beyond-supercohomology anomalies cannot be realized by fermionic topological orders, and Cheng--Wang--Yang~\cite{Cheng:2024awi} argued using a \emph{spatial} rotation symmetry that all in-supercohomology anomalies can be realized by fermionic topological orders. Thus those parts of cases~\ref{case1}--\ref{case4} are not new. However, the IR phases of these theories are not well studied, except for adjoint QCD, and that aspect of our analysis in cases~\ref{case2}--\ref{case4} is new. Case~\ref{case5} does not have a $\Z_{2n}^F$ symmetry and, to our knowledge, is completely new.



\textit{Note:} Recent independent work by Wan and Wang~\cite{Wan:2025ymd} overlaps with ours. They use the Atiyah--Patodi--Singer $\eta$-invariants to analyze discrete Abelian symmetries of Weyl fermions and determine minimal extension kernels in a range of explicit examples, with particular emphasis on Standard Model applications. In a follow-up paper with Yau~\cite{ZJSNew}, they use this result to provide a possible explanation of the three families in the Standard Model. Our approach instead starts from extended supercohomology and fusion 2-categories, proves a general existence/no-go dichotomy for finite symmetries, and then extracts physical consequences. Our results agree where they overlap; see Table~\ref{tab:results_TQFT} and \Cref{thm:gapless}.

\section{Fermionic symmetries and anomalies}\label{sec:fermionic-symmetries-anomalies}
Supercohomology is an invariant of fermionic symmetries in the same way that ordinary cohomology is for bosonic symmetries. Thus, before we introduce supercohomology, we review the data that enter into a fermionic symmetry.

A fermionic system can be characterized by its symmetry action, which forms a fermionic symmetry group $G_f$. Beyond the group structure of $G_f$, two additional pieces of data are important: (1) a map $\rho\colon G_f\rightarrow \Z_2$ such that the symmetry element is antiunitary or unitary if the image under $\rho$ is $1$ or $0$, respectively, and (2) a central $\Z_2$ subgroup $\langle (-1)^F\rangle \subset G_f$ in the kernel of $\rho$ generated by fermion parity. This motivates describing the fermionic symmetry using the following three pieces of data~\cite[\S 7]{Ben88}:\footnote{See also~\cite[\S 2.1]{Ste22} for more on the connection between these data and fermionic symmetries.}
\begin{itemize}
\item a (bosonic) symmetry group $G\coloneqq G_f/\langle(-1)^F\rangle$ acting on the bosonic degrees of freedom;
\item a class $s \in \rH^1(G;\Z_2)$, corresponding to $\rho$;
\item a class $\omega \in \rH^2(G; \Z_2)$, classifying the extension of $G$ by the $\Z_2$ subgroup $\langle(-1)^F\rangle$ to get $G_f$.
\end{itemize}
Throughout the paper, we sometimes use the bosonic symmetry group $G$ to refer to a fermionic symmetry, with its associated $s$ and $\omega$ twists implicit.

When referring to fermionic symmetry groups, we will use a superscript $F$ to indicate that part of the symmetry is generated by fermion parity $(-1)^F$.
\begin{example}
\label{spincn}
$\Z_{2n}^F$ refers to the fermionic symmetry group with $G_f=\Z_{2n}$ and $G=\Z_n$, for which the quotient map $G_f\to G$ is reduction modulo $n$, $\Z_{2n}\to\Z_n$. This symmetry is purely unitary.

In terms of $s$ and $\omega$, $s = 0$ in $\rH^1(\Z_n;\Z_2)$. If $n$ is even, $\rH^2(\Z_n;\Z_2)\cong\Z_2$ and $\omega$ is the unique nonzero element; if $n$ is odd, then $\Z_{2n}^F\cong\Z_n\times\Z_2^F$, so $\omega = 0$.
\end{example}

Later, when we discuss supercohomology, it will be an invariant of the entire fermionic symmetry group, including $s$ and $\omega$.

\subsection{Supercohomology and beyond-supercohomology anomalies}\label{sec:supercohomology-anomalies}
We begin with a lightning review of supercohomology\footnote{Here and throughout, \emph{supercohomology} means extended
supercohomology in the sense of
Refs.~\cite{Wang:2017moj,KT17,Gaiotto:2017zba,D11}; equivalently, the
generalized cohomology theory associated to the Picard 2-groupoid
$\tsVect^\times$.} and explain why it replaces ordinary cohomology, which classifies bosonic anomalies, in fermionic settings. Quantum anomalies are fundamentally linked to projective representations \cite{2023arXiv230708147F}. Given a symmetry group $G$, a projective unitary representation of $G$ is defined to be a function $\rho\colon G\rightarrow\U(n)$ for some positive integer $n$, which is a homomorphism up to a ``correction term'' $\omega(g,h)\in\U(1)$, in that for any $g,h\in G$ we have:
\beq
\rho(g)\cdot \rho(h) = \omega(g, h)\rho(gh).
\eeq
Associativity of the representation operators requires
\begin{equation}
    \omega(g,h)\omega(gh,k)=\omega(h,k)\omega(g,hk),
\end{equation}
so $\omega$ is a $\U(1)$-valued 2-cocycle.
Physically, we allow the presence of such nontrivial $\omega(g, h)$ taking values in $\U(1)$ because a vector in the physical Hilbert space represents the same physical state if it is multiplied by a $\U(1)$ factor, so that the action of $G$ on physical states is well-defined in a projective unitary representation. This explains the origin of the $\U(1)$ coefficient in the classification of in-cohomology anomalies in bosonic systems. In particular, anomalies for $G$-symmetries in (0+1)d quantum mechanical systems are exactly in one-to-one correspondence with projective representations of $G$, and are classified by $\rH^2(G;\U(1))$.\footnote{In this paper, when we write $\rH^n(G; A)$, we always mean the cohomology of the classifying space $BG$: $\rH^n(BG; A)$. This also applies to supercohomology in Sec.~\ref{sec:supercohomology}.} In $n$ spatial dimensions with $n\geq 1$, various $G$-defects may carry nontrivial projective representations of $G$, and the associated in-cohomology anomalies are classified by $\rH^{n+2}(G;\U(1))$ \cite{Chen:2011pg,2013PhRvD..88d5013W}.

In fermionic systems, the presence of fermion parity and local fermion excitations permits additional actions. In the context of 3-dimensional fermionic topological orders, in addition to multiplication by $\U(1)$ on the physical Hilbert space, there are local fermion excitations and their ``condensations,'' which physically correspond to the Kitaev chain.
In categorical language~\cite{D11}, these ``symmetry actions'' are encoded by the Picard 2-groupoid $\tsVect^\times$, which forms a nontrivial 3-group~\cite{Benini:2018reh}, with the 0-form part $\U(1)$, 1-form part $\Z_2$, and 2-form part $\Z_2$. Therefore, roughly speaking, we should replace $\omega(g,h)$ with actions in the 3-group \cite{Gaiotto:2017zba,Yu:2020twi}, as explained in more detail in Sec.~\ref{sec:supercohomology}. The resulting \textit{generalized cohomology} theory is precisely \textit{supercohomology} $\SH^*(G_f)$.\footnote{We use the indexing convention that $\SH^{n+2}(G_f)$ classifies the in-supercohomology anomalies of fermionic QFTs with $G_f$ symmetry in $n$ spatial dimensions.} Thus, we should anticipate that supercohomology in the fermionic setting plays a role analogous to ordinary cohomology in the bosonic setting.

We want to emphasize that supercohomology is a cohomology theory of fermionic symmetry groups. In the mathematical physics literature, it is more common to phrase this as the \term{twisted supercohomology} of $G$, as in~\cite{D11}, rather than the supercohomology of $G_f$. This perspective generalizes the use of the twisted ordinary cohomology groups $\rH^*(G; \U(1)_\rho)$ to classify projective representations and anomalies of a bosonic symmetry $G$ with $\rho\colon G\to\Z_2$ specifying whether elements of $G$ act unitarily or antiunitarily. In the fermionic case, both pieces of data $(s,\omega)$ are used to define the twist.

However, in the bosonic and fermionic contexts, some 't Hooft anomalies are not described by cohomology and supercohomology classes, respectively.
We review the classification of 't Hooft anomalies in fermionic settings in Sec.~\ref{sec:thooft-anomalies}.
We call these additional bosonic and fermionic anomalies \textit{beyond-cohomology} and \textit{beyond-supercohomology} anomalies, respectively. We will discuss the physical meaning of beyond-supercohomology anomalies in Sec.~\ref{sec:gaplessness}.

\subsection{Fermionic symmetries and 't Hooft anomalies}\label{sec:thooft-anomalies}

We next clarify the mathematical formulation of fermionic symmetries and the classification of their associated 't Hooft anomalies.

\label{sec:supercohomology}
According to the mechanism of anomaly inflow, possible quantum anomalies associated with $G_f$ or $G$ are represented by fermionic symmetry-protected topological (SPT) phases in one dimension higher. In the decorated domain wall picture, these SPT phases are constructed by decorating symmetry domain walls with lower-dimensional invertible states~\cite{Kit13,Kit15,Wang:2017moj,PhysRevX.10.031055}. An invertible state is a short-range-entangled gapped phase with an inverse under stacking, such that stacking the two states yields a trivial gapped vacuum. In physically relevant dimensions, this construction yields four layers of data:
\begin{itemize}
\item the $p+ip$ layer, originating from decoration by the $p+ip$ topological superconductor;
\item the Majorana layer, originating from decoration by the Kitaev chain;
\item the Gu--Wen layer, originating from decoration by complex fermions; and
\item the Dijkgraaf--Witten layer, originating from decoration by bosonic SPTs protected by $G$ symmetry.
\end{itemize}
Further mathematical details of this layered construction are provided in Refs.~\cite{PhysRevX.10.031055,2024PhRvB.110w5117R,2025arXiv251225069N,Debray:2025kfg}. In $n$ spatial dimensions, the four layers together form the group of SPTs protected by a fermionic symmetry group $G_f$, which we denote by $\mho^{n+1}(G_f)$.\footnote{We use Freed--Hopkins'~\cite{FH} mathematical model for the group of SPTs $\mho^n(G_f)$, which is the abelian group of deformation classes of $n$-dimensional reflection-positive invertible field theories on manifolds with a $(BG, s, \omega)$-twisted spin structure. See~\cite{FH, Fre19} for additional discussion of this model and its relation to SPTs and~\cite{FH, Cam17, KT17, BC18, Wang:2017moj, Gaiotto:2017zba, FH20, ABK21, Deb21, BCHM22, BottSpiral} for computational evidence supporting this model.
Just as for supercohomology, one can either think of $\mho^*(G_f)$ as an invariant of a fermionic group $G_f$ or as a twisted generalized cohomology group associated to the bosonic group $G$ and the twisting data $(s,\omega)$. This parallels the classification of bosonic SPTs in the presence of antiunitary symmetries.}

Omitting decorations by the $p+ip$ invertible states in 2 spatial dimensions leaves the Majorana, Gu--Wen, and Dijkgraaf--Witten layers. These remaining layers form the subgroup of SPTs in the supercohomology of $G_f$, denoted by $\SH^{n+1}(G_f)$. The $p+ip$ layer is classified by $\rH^{n-2}(G;\Z)$, where $G = G_f/\langle \Z_2^F\rangle$ is the bosonic symmetry associated to $G_f$. For quantum anomalies in 3 spatial dimensions, $n=4$, so this group is $\rH^2(G;\Z)$. Since $\rH^2(G;\Z)\cong \rH^1(G;\U(1))$, elements of the $p+ip$ layer are canonically associated to homomorphisms $\rho\colon G\rightarrow \U(1)$.

From the point of view of category theory, the invertible states underlying supercohomology are organized as the Picard 2-groupoid $\tsVect^\times$. The objects of $\tsVect^\times$ are Morita-invertible complex superalgebras, the morphisms are super Morita equivalences, and the 2-morphisms are intertwiners \cite[Section 2.2]{JF2}. Within this framework, the only nontrivial object up to Morita equivalence is the Clifford algebra $\mathrm{Cl}(1)$. The resulting homotopy groups and $k$-invariants reproduce the nontrivial 3-group structure summarized in Sec.~\ref{sec:supercohomology-anomalies}.\footnote{See~\cite{GJOS17} for a comprehensive discussion of these data and how to calculate them for a Picard $2$-groupoid, as well as~\cite[\S 4.3.5]{DG18} for $\tsVect^\times$ specifically.}

This perspective on $\tsVect^\times$ naturally connects supercohomology to the fusion 2-category description of fermionic topological orders. In particular, a braided fusion $2$-category of operators is expected to be associated with each topological order in three spatial dimensions~\cite[Section 2.2]{Decoppet:2025eic}. In the bosonic setting, this braided fusion 2-category is \emph{non-degenerate}, meaning that only trivial operators are transparent. Mathematically, this non-degeneracy means that the \emph{sylleptic center} \cite{Cr} of the braided fusion 2-category is equivalent to $\tVect$. In the fermionic setting, a local fermion can braid trivially with all other operators. A braided fusion 2-category whose sylleptic center is equivalent to $\tsVect$ is therefore called a non-degenerate fermionic braided fusion 2-category.

To enrich a non-degenerate fermionic braided fusion 2-category $\fB$ with a $G$ symmetry, we introduce $G$-defects that permute the theory's operators. Mathematically, the symmetry action involves a nontrivial map from $G$ to the Picard groupoid of $\fB$, and an obstruction to this map is captured by a 5-groupoid containing $\tsVect^\times$. This strongly suggests that anomalies for fermionic topological orders take values in supercohomology. Further mathematical details, including the categorical interpretation of the symmetry extension construction, are presented in Refs.~\cite{Decoppet:2025eic,D8,D11,Debray:2025kfg}.

In addition, there is a natural homomorphism $i\colon \SH^{n+1}(G_f)\to\mho^{n+1}(G_f)$, which one can think of as sending an in-supercohomology SPT to its image in the group of all $(n+1)$-dimensional $G_f$-symmetric fermionic SPTs. There is another map $p\colon \mho^{n+1}(G_f)\to H^{n-2}(G;\Z_s)$, which is the data of the $p+ip$ layer of an SPT.\footnote{Here $\Z_s$ denotes the local system in which $g\in G$ acts on $\Z$ through the homomorphism $s\colon G\to\{\pm 1\}$.} Thus we have a sequence
\beq\label{eq:ses_anomaly}
\SH^{n+1}(G_f) \xrightarrow{i} \mho^{n+1}(G_f) \xrightarrow{p} \rH^{n-2}(G;\Z_s),
\eeq
which is exact in the middle entry. 
By anomaly inflow, the anomaly of a specific theory is represented by an element of $\mho^{n+1}(G_f)$; it is a supercohomology anomaly if it lies in the image of $i$, and a beyond-supercohomology anomaly if its image under $p$ is nontrivial.

As a concrete example, consider $G_f = \Z_4^F$ (see \cref{spincn}), in which the square of a generator is the fermion-parity element $(-1)^F$. The 't Hooft anomaly of a theory with a $\Z_4^F$ symmetry is a 4-dimensional SPT with the same symmetry. Such SPTs are classified by $\mho^5(\Z_4^F)\cong \Z_{16}$~\cite[\S 2]{Gia73a}. The corresponding supercohomology group is $\SH^5(\Z_4^F)\cong\Z_8$~\cite[Proposition E.20]{Debray:2025kfg}, the even elements of the $\Z_{16}$ of SPTs. In contrast, the odd elements of $\Z_{16}$ are the beyond-supercohomology anomalies. We can embed this symmetry into the $\U(1)^F$ symmetry and think of it as part of a continuous chiral symmetry. This symmetry is discussed in Refs.~\cite{CO2,Cheng:2024awi,Debray:2025kfg,Hsi18,Wan:2025lad,Garcia-Etxebarria:2018ajm}.

As another example, consider $G_f=\Z_4\times \Z_2^F$, where the $\Z_4$ factor is independent of fermion parity. In 3 spatial dimensions, the 't Hooft anomaly is an element of $\mho^5(\Z_4\times\Z_2^F)\cong \Z_{4}$~\cite[Example 7.3.3]{BG10}, and $\SH^5(\Z_4\times\Z_2^F)\cong\Z_2$~\cite[Proposition E.16]{Debray:2025kfg}. Once again the map $i\colon\SH^5(\Z_4\times\Z_2^F)\to\mho^5(\Z_4\times\Z_2^F)$ is the inclusion of the even elements of $\Z_4$. The odd elements of $\Z_{4}$ are the beyond-supercohomology anomalies. In Sec.~\ref{sec:applications}, we discuss a Weyl fermion coupled to a bosonic field with $\Z_4\times \Z_2^F$ symmetry whose anomaly realizes one of the beyond-supercohomology SPTs. This symmetry is also discussed in Refs.~\cite{Wan:2025lad,Hsi18,Garcia-Etxebarria:2018ajm}.

For each fermionic symmetry group $G_f$ considered in Table~\ref{tab:results_TQFT}, we give $\SH^5(G_f)$ and $\mho^5(G_f)$ in Table~\ref{tab:classification}. These groups were computed in Refs.~\cite{Hsi18,PhysRevX.10.031055,Debray:2025kfg}.

\begin{table}[!htbp]
\centering
\renewcommand{\arraystretch}{1.3}
\begin{tabular}
{c | c | c}
\hline \hline 
$G_f$ & $\SH^5(G_f)$ & $\mho^5(G_f)$\\ \hline\hline
$\Z_{3^k}\times \Z_2^F$ & $(\Z_{3^{k}},\text{DW})$ & $\Z_{3^{k-1}}\oplus \Z_{3^{k+1}}$\\ \hline 
$\Z_{p^k}\times \Z_2^F,p\geq 5$ & $(\Z_{p^{k}},\text{DW})$ & $\Z_{p^{k}}\oplus\Z_{p^{k}}$\\ \hline 
$\Z_2\times \Z_2^F$ &  trivial & trivial\\ \hline 
{$\Z_{2^k}\times \Z_2^F,k\geq 2$} & $(\Z_{2^{k-1}},\text{DW})$ & $\Z_{2^{k-2}}\oplus \Z_{2^{k}}$ \\ \hline 
$\Z_4^F$ &  $(\Z_8,\text{Maj})$ & $\Z_{16}$\\ \hline 
$\Z_{2^{k+1}}^F,k\geq 2$ & $(\Z_{2^{k+1}},\text{GW})\oplus \,({\Z_2},\text{Maj})$ & $\Z_{2^{k+3}}\oplus \Z_{2^{k - 1}}$ \\ 
\hline 
$\Z_{2^{k+1}}^F\times \Z_2^T,k\geq 2$ & $\textcolor{Orange}{\Z_2} \oplus (\Z_2,\text{DW})\oplus(\Z_{4},\text{GW})$ &  $\Z_2\oplus\Z_{4}$ \\
\hline  
\hline
\end{tabular}
\caption{Fermionic symmetries and their degree-$5$ supercohomology and SPT groups: $\SH^5(G_f)$ classifies in-supercohomology anomalies, while $\mho^5(G_f)$ classifies all anomalies of fermionic QFTs with $G_f$ symmetry in 3 spatial dimensions. For supercohomology, we indicate the layer of each generator: Majorana (Maj), Gu--Wen (GW), or Dijkgraaf--Witten (DW). The orange generator has zero image in the group of SPTs under the map in Eq.~\eqref{eq:ses_anomaly}.}
\label{tab:classification}
\end{table}

\section{Construction of fermionic gauge theories and candidate states in the IR}\label{sec:construction}
\subsection{Construction of fermionic gauge theories}\label{sec:construction-gauge-theories}
Nontrivial gapped states usually take the form of finite gauge theories \cite{Kitaev:2005hzj,Lan_2019,JF,Decoppet:2025eic}, and in 3 spatial dimensions these are the only possibilities. This motivates the symmetry extension construction, which proceeds as follows in the bosonic case. We start with a finite symmetry group $G$, and consider the symmetry extension of $G$ according to a short exact sequence
\beq\label{eq:SES}
1\rightarrow K \xrightarrow{j} H \xrightarrow{p} G\rightarrow 1,
\eeq
such that for a given element $\alpha \in \rH^{n+2}(G;\U(1))$, the pullback along $p$ trivializes $\alpha$ in the sense that $p^*\alpha = 0 \in \rH^{n+2}(H;\U(1))$. Thus, we can gauge the subgroup $K$ in an $H$-symmetric state to obtain a $K$-gauge theory. The anomalous symmetry action of $G$ on the resulting $K$-gauge theory, a gapped topological order, then gives the prescribed anomaly $\alpha$.

The symmetry extension construction in the bosonic setting is supported by path-integral formulations and lattice constructions \cite{Wang:2017loc}. However, generalizations of these constructions to the fermionic setting are much harder. In Ref.~\cite{Debray:2025kfg}, we interpret the construction in the context of fusion 2-categories, obtaining a rigorous generalization for unitary symmetries.\footnote{For the time-reversal example in Table~\ref{tab:results_TQFT}, we only compute the pullback of the relevant twisted supercohomology classes. Interpreting this computation as a construction of a time-reversal-symmetric gauge theory would require a categorical framework for antiunitary symmetries, which is still under development; we therefore treat this example as formal.} The central difference is that the ``coefficient'' is a $\U(1)$ action in the bosonic setting but a 3-group action in the fermionic setting, as explained in Sec.~\ref{sec:supercohomology}.

Formally, given a finite fermionic symmetry group $G_f$,\footnote{As reviewed in Sec.~\ref{sec:thooft-anomalies}, we have two ways to encode the symmetry group of a fermionic symmetry, either in terms of the bosonic symmetry group or the fermionic symmetry group. We will use $G$ to denote the bosonic symmetry group, and use $G_f$ with subscript $f$ to denote the fermionic symmetry group.} we just need to replace ordinary cohomology with supercohomology. Given a short exact sequence of bosonic symmetry groups of the form~\eqref{eq:SES}, there is a unique lift to fermionic symmetry groups: if $\omega\in \rH^2(G;\Z_2)$ classifies the double cover $G_f\to G$, let $H_f\to H$ be classified by $p^*\omega$. We call the fermionic symmetry with the data $(H, p^*s, p^*\omega)$ the \term{pullback fermionic symmetry}. Then $j^*p^*\omega = 0$, so $K$ does not mix with fermion parity.

We can now proceed as in the bosonic setting. For a specified element $\alpha \in \SH^{n+2}(G_f)$, if the pullback along $p$ in Eq.~\eqref{eq:SES} trivializes $\alpha$ in the sense that $p^*\alpha = 0 \in \SH^{n+2}(H_f)$, we can gauge the subgroup $K$ in an $H_f$-symmetric state to obtain a $K$-gauge theory with an anomalous $G_f$-action corresponding to $\alpha$.

Ref.~\cite{Tachikawa:2017gyf} showed that, given any in-cohomology anomaly $\alpha\in \rH^{n+2}(G;\U(1))$ of bosonic systems with \textit{finite} $G$, it is always possible to obtain a surjective homomorphism $p\colon H\twoheadrightarrow G$ from a finite group $H$ such that $p^*\alpha = 0$. In Ref.~\cite{Debray:2025kfg}, we prove that this is also true for $\alpha\in\SH^5(G_f)$, leading to the following result.
\begin{thm}\label{thm:always_triv}
In 3 spatial dimensions, every quantum anomaly of a finite fermionic symmetry that is captured by supercohomology can be realized by a fermionic topological order using the symmetry extension construction.
\end{thm}

The proof proceeds by unfolding the supercohomology anomaly into three different layers, each of which is captured by ordinary cohomology. Then we just need to trivialize one layer at a time. We give a sketch of the proof in Sec.~\ref{sec:proof-theorem-1}, leaving the full mathematical details for Ref.~\cite{Debray:2025kfg}.

\subsection{Proof of \cref{thm:always_triv}}\label{sec:proof-theorem-1}

We now summarize the formal trivialization result underlying \cref{thm:always_triv}; it holds for arbitrary $s$, although its categorical gauge-theory interpretation presently requires $s=0$.
Let $G_f$ be a finite fermionic symmetry group with bosonic quotient $G$, and let $\alpha\in\SH^n(G_f)$ with $n\geq 5$. The formal statement of the result is that there is an algorithmic construction of a finite group $\widetilde G$ and a surjective homomorphism $\rho\colon \widetilde G\twoheadrightarrow G$ such that \begin{equation}
    \rho^*(\alpha) = 0\in \SH^n(\widetilde G_f),
\end{equation}
where, as in Sec.~\ref{sec:construction-gauge-theories}, $\widetilde G_f$ denotes the pullback fermionic symmetry with data $(\widetilde G,\rho^*s,\rho^*\omega)$.

Writing $K\coloneqq\ker(\rho)$, we obtain a short exact sequence of groups
\begin{equation}
1 \rightarrow K \rightarrow \widetilde{G}\xrightarrow{\rho} G \rightarrow 1.
\end{equation}
For $s=0$, gauging the $K$ symmetry of a $\widetilde{G}_f$-symmetric state produces a $K$-gauge theory with the prescribed anomaly $\alpha$; for $s\ne0$, the construction establishes only the formal pullback trivialization.


We employ the following lemma \cite[Theorem 4]{DLM24} to guarantee that it is possible to trivialize each layer; its algorithmic refinement is stated in Ref.~\cite[Theorem 3.5]{Debray:2025kfg}, and versions or special cases appear in Refs.~\cite[\S V.A]{Wang:2017loc}, \cite[\S 2.7]{Tachikawa:2017gyf}, and \cite[Proposition 4.2.3]{D9}.
\begin{lem}\label{thm:folk}
    Let $G$ be a finite group, $M$ be a $\Z[G]$-module, $n\ge 3$, and $\alpha\in \rH^n(G; M)$. Then there is an algorithmic construction of a finite group $\widetilde G$ and a surjective homomorphism $\rho\colon \widetilde G\twoheadrightarrow G$ such that $\rho^*(\alpha) = 0\in\rH^n(\widetilde G;\rho^*M)$.
\end{lem}


The proof proceeds by trivializing the anomaly one layer at a time. First, the map
\begin{equation}
(\bl)_{\mathrm{Maj}}\colon \SH^n(G_f)\longrightarrow \rH^{n-2}(G;\Z_2)
\end{equation}
sends a class $\alpha$ to its Majorana layer $\alpha_{\mathrm{Maj}}$; by \Cref{thm:folk}, there is a finite extension $p_1\colon G_1\twoheadrightarrow G$ on which this layer becomes trivial. Next, using reduced supercohomology, which retains the remaining Gu--Wen and Dijkgraaf--Witten layers \cite{Wang:2017moj}, another application of \Cref{thm:folk} yields a finite extension $p_2\colon G_2\twoheadrightarrow G_1$ that trivializes the Gu--Wen layer of $p_1^*\alpha$. Finally, \Cref{thm:folk} provides a finite extension $\widetilde G\twoheadrightarrow G_2$ that trivializes the remaining Dijkgraaf--Witten layer of $p_2^*p_1^*\alpha$. Thus, the full supercohomology anomaly vanishes after pullback to $\widetilde G_f$, where $(G_1)_f$, $(G_2)_f$, and $\widetilde G_f$ denote the corresponding pullback fermionic symmetries.

\subsection{Candidate states in the IR}\label{sec:candidate-states}
Building on this general framework, we now examine specific symmetries and anomalies relevant to physical contexts, identifying fermionic topological orders that realize each anomaly and determining when such a realization is impossible. We focus on cyclic symmetry groups, including $G_f = \Z_{2n}^F$ (\cref{spincn}), that appear frequently in physics, and make use of supercohomology computations from~\cite{Debray:2025kfg} and $\mho^*(G_f)$ computations from~\cite{Hsi18,PhysRevX.10.031055}. We are interested in theories in $3$ spatial dimensions, and hence we use $\SH^5$ and $\mho^5$.

The general procedure is straightforward:
\begin{enumerate}
    \item For each symmetry group $G_f$, calculate $\SH^5(G_f)$ and $\mho^5(G_f)$.
    \item For each element $\alpha$ of a generating set of $\SH^5(G_f)$, calculate a minimal\footnote{Here, ``minimal'' means that $\# H/\# G$ is as small as possible among those for which the pullback of $\alpha$ to $H$ is $0$.} group $H$ with a surjective map $p\colon H\to G$ such that $p^*\alpha = 0$. Let $K\coloneqq\ker(p)$; then the gauge theory that realizes the anomaly $\alpha$ is the finite $K$-gauge theory with gauge group $K$ given by the subgroup of $H$ in Eq.~\eqref{eq:SES}.
\end{enumerate}

The results are summarized in Table~\ref{tab:results_TQFT}. We will comment on the potential applications of these results in physics in Sec.~\ref{sec:applications}.

\begin{table}[!htbp]
\centering
\renewcommand{\arraystretch}{1.3}
\begin{tabular}
{c | c | c}
\hline \hline 
$G_f$ & $\SH^5(G_f)$ & $K$ (gauge group) \\ \hline\hline
$\Z_{p^k}\times \Z_2^F$ & $(\Z_{p^{k}},\text{DW})$ & $\Z_{p^m},m=\lceil \frac{k}{2}\rceil$\\ \hline 
{$\Z_{2^k}\times \Z_2^F$} & $(\Z_{2^{k-1}},\text{DW})$ & $\Z_{2^m},m=\lceil \frac{k-1}{2}\rceil$ \\ \hline 
$\Z_4^F$ &  $(\Z_8,\text{Maj})$ & $\Z_4$\\ \hline 
\multirow{2}{*}{$\Z_{2^{k+1}}^F,k\geq 2$} & $(\Z_{2^{k+1}},\text{GW})$ & $\Z_{2^m},m=\lceil \frac{k}{2}\rceil$\\ 
&  $\oplus \,({\Z_2},\text{Maj})$ & $\Z_4$ \\
\hline 
\multirow{2}{*}{$\Z_{2^{k+1}}^F\times \Z_2^T,k\geq 2$} & $\textcolor{orange}{\Z_2} \oplus (\Z_2,\text{DW})$ & $\Z_2$ \\
& $\oplus(\Z_{4},\text{GW})$ & $\Z_4\times \Z_2$\\
\hline  
\hline
\end{tabular}
\caption{Fermionic symmetries, the corresponding groups of anomalies in supercohomology in 3 spatial dimensions, and the kernels $K$ of extensions trivializing the corresponding anomaly classes. The second column decomposes these groups into direct summands and indicates the layer of each generator: Majorana (Maj), Gu--Wen (GW), or Dijkgraaf--Witten (DW) (see Sec.~\ref{sec:thooft-anomalies}). Orange generators have zero image in the group of full fermionic anomalies. The third column gives the candidate states in terms of $K$-gauge theories with a finite gauge group $K$. In the first row, $p$ is an odd prime.}
\label{tab:results_TQFT}
\end{table}

\FloatBarrier

\section{Symmetry-Enforced Gaplessness}\label{sec:gaplessness}
\subsection{Physical picture}\label{sec:gaplessness-physical-picture}
Not every anomaly of a fermionic QFT can be described by supercohomology, analogously to the limitations of ordinary cohomology in classifying anomalies of bosonic QFTs. Thus, it is interesting to ask whether symmetric gapped states can capture these additional anomalies. The fusion 2-category picture already suggests that the answer is negative. 

\begin{thm}\label{thm:gapless}
In 3 spatial dimensions, let $\alpha$ be a quantum anomaly of a finite fermionic symmetry that cannot be captured by supercohomology. Then $\alpha$ cannot be realized by a symmetry-preserving fermionic gapped state.
\end{thm}

We now give a heuristic explanation of the mechanism underlying \cref{thm:gapless}. Consider the $\U(1)^F$ symmetry of a single (3+1)-dimensional left-handed Weyl fermion $\chi$~\cite{peskin2018introduction},
\beq\label{eq:Lagrangian_Weyl}
\mathcal{L} = i\bar{\chi}\slashed{\partial} \chi,
\eeq
where $e^{i\theta}\in \U(1)^F$ acts on $\chi$ by $\chi\rightarrow e^{i\theta}\chi$.
The group of possible anomalies is $\mho^5(\U(1)^F)\cong\Z\oplus\Z$. In triangle-diagram language \cite{peskin2018introduction}, let $a_3$ and $a_1$ denote the cubic $\U(1)^F$ and mixed $\U(1)^F$--gravitational coefficients, respectively. For the odd fermion charges appropriate to $\U(1)^F$, $a_3-a_1=24k$ with $k\in\Z$, so $(a_1,k)$, rather than $(a_1,a_3)$, gives integral coordinates on the anomaly lattice \cite[Sec.~II.2]{Wan:2025lad}. Compactification on K3 identifies the $p+ip$-layer projection with $a_1$ \cite[Sec.~3.2]{Debray:2025kfg}, so its kernel---the $k$ direction---is captured by supercohomology, while a nonzero $a_1$ lies beyond supercohomology. A charge-$1$ left-handed Weyl fermion has $(a_1,k)=(1,0)$, although its raw cubic and mixed-gravitational coefficients are both nonzero. We detect its beyond-supercohomology component by analyzing the nontrivial $\U(1)$ vortex. Upon dimensional reduction, the vortex core hosts a (1+1)-dimensional chiral fermion and therefore has a gravitational anomaly. This detects the nonzero beyond-supercohomology component of the parent (3+1)-dimensional Weyl-fermion anomaly.\footnote{This analysis is reminiscent of the Smith homomorphism connecting anomalies on a defect with anomalies of the whole system \cite{COSY20,Hason:2020yqf,Debray:2023ior,Debray:2025iqs}.}

For anomalies of unitary finite symmetries in (3+1)d fermionic systems, the extra piece of data that lies beyond supercohomology is captured by a homomorphism $\rho\colon G\rightarrow \U(1)$ \cite{PhysRevX.10.031055} (see Sec.~\ref{sec:thooft-anomalies}). To understand the physical meaning of the homomorphism $\rho$, we also construct a $G$-vortex such that $G$ symmetry acts on the two transverse spatial dimensions according to the homomorphism $\rho$. Then the beyond-supercohomology anomaly is captured by the fact that this $G$-vortex carries a nontrivial gravitational anomaly. However, the gravitational anomaly in (1+1)d suggests that the core of the $G$-vortex is always gapless. This gives a heuristic argument for why, in the presence of beyond-supercohomology anomalies, there is always a gapless mode.

We provide a mathematical proof in Sec.~\ref{sec:proof-theorem-2}, following the cyclic-symmetry argument of Ref.~\cite{CO2} and its generalization in Ref.~\cite[Theorem 3.1(2)]{Debray:2025kfg}. This gives a comprehensive argument for \cref{thm:gapless} from both physical and mathematical perspectives.

\subsection{Proof of \cref{thm:gapless}}\label{sec:proof-theorem-2}

We now give a summary of our proof of \cref{thm:gapless}, first for unitary symmetries, so that $s=0$. As reviewed in Sec.~\ref{sec:thooft-anomalies}, the anomaly of a fermionic QFT with fermionic symmetry group $G_f$ is represented by a class $\alpha\in \mho^5(G_f)$, and there is a natural projection
\begin{equation}
(\bl)_{p+ip}\colon \mho^5(G_f) \rightarrow \rH^2(G;\Z)\cong \rH^1(G;\U(1)),
\end{equation}
where as before $G \coloneqq G_f/\langle \Z_2^F\rangle$ is the bosonic symmetry.
From the decorated domain wall point of view on SPTs (see Sec.~\ref{sec:thooft-anomalies}) $\alpha_{p+ip}\in\rH^1(G;\U(1))$ describes decorations with the $p+ip$ topological superconductor. The data of $\alpha_{p+ip}$ is also equivalent to a homomorphism $\alpha_{p+ip}\colon G\rightarrow\U(1)$. As shown in~\cite{Hsi18,Cheng:2024awi}, given any nonzero element $\alpha_{p+ip}\in \rH^1(G;\U(1))$, it is impossible to trivialize this element by a short exact sequence
\begin{equation}
1 \rightarrow K \rightarrow \widetilde{G}\rightarrow G \rightarrow 1,
\end{equation}
such that $K$ is a finite group,
unlike what we have done for supercohomology. Thus the layered approach to symmetry extension we used in Sec.~\ref{sec:proof-theorem-1} does not work for any beyond-supercohomology anomaly $\alpha$: by the exact sequence~\eqref{eq:ses_anomaly}, $\alpha$ is beyond-supercohomology if and only if $\alpha_{p+ip}\ne 0$. However, this does not yet imply that there is no topological order with anomaly $\alpha$, as it is possible to construct anomalous topological orders using methods other than symmetry extension. Thus, we had to take a different approach, based on the approach in~\cite{CO2}, to show that such $\alpha$ cannot be realized.
Specifically, we analyze the partition function of the invertible field theory $\alpha$ corresponding to the anomaly SPT on $\mathrm{K3} \times S^1$. From the homotopical classification of invertible field theories~\cite{FH}, one can calculate $\alpha(\mathrm{K3}\times S^1)$ using methods from algebraic topology.
Unlike the original approach in~\cite{CO2}, our method does not rely on the Atiyah--Singer index theorem and is therefore generalizable to all fermionic systems (field theories). Following the analysis of the partition function on the K3 surface established in Ref.~\cite{CO2}, we complete the proof.

The key lemma in our proof is as follows. For a finite fermionic symmetry group $G_f$ and $g\in G$, let $S_g^1$ denote a circle equipped with the bounding spin structure and a principal $G$-bundle with monodromy $g$.
\begin{lem}\label{thm:folk2} 
    Let $G_f$ be a finite group with $s=0$, and let $\alpha\in \mho^5(G_f)$. Suppose that $\alpha_{p+ip}\ne 0$ in $\rH^1(G;\U(1))$, i.e.\ $\alpha$ is a beyond-supercohomology SPT.
    Consider the invertible field theory representing $\alpha$ under anomaly inflow~\cite{FH}. 
    Its partition function on $\mathrm{K3}\times S^1_g$ is given by $\alpha_{p+ip}(g)\in \U(1)$.
\end{lem}
We make this calculation using a mathematical model for ``compactification on the K3 surface'', viewed as dimensional reduction of a field theory along a closed manifold \cite[\S 9.2]{SchommerPries:2018}.\footnote{See also~\cite{YY23} for further discussion.} The outcome of the compactification is an invertible field theory in one dimension, which is exactly characterized by $\alpha_{p+ip} \in \mho^1(G_f)$. Evaluating the partition function of $\alpha_{p+ip}$ on $S^1_g$ then gives the desired result. The mapping-torus trace of the $g$-action uses the nonbounding spin circle, but its difference from the bounding choice lies in the image of $\Omega_1^\Spin$; after taking the product with K3, the difference lies in $\Omega_5^\Spin=0$, so the two evaluations agree \cite[\S 3.2]{Debray:2025kfg}.

The rest of the proof follows the analysis in Ref.~\cite{CO2}, and we briefly repeat it here for the reader's convenience. On the one hand, Lemma~\ref{thm:folk2} implies that for \emph{any} quantum theory in 3 spatial dimensions with anomaly $\alpha$, either $Z[\mathrm{K3}]=0$ or there is some $g\in G$ with $\alpha_{p+ip}(g)\ne1$ that acts nontrivially on $Z[\mathrm{K3}]$. Symmetry preservation requires $Z[\mathrm{K3}]$ to be $G$-invariant; this is the step used for cyclic symmetries in Ref.~\cite{CO2} and for general finite symmetries in Ref.~\cite[Sec.~3.2]{Debray:2025kfg}. Combining these two facts, $Z[\mathrm{K3}] = 0$.
On the other hand, a topological order is described by a (3+1)-dimensional unitary TQFT $Z$ which assigns a one-dimensional Hilbert space to $S^3$. We recall the argument of Ref.~\cite{CO2} that $Z[\mathrm{K3}]\ne0$. Unitarity gives $Z[S^4]>0$, and Ref.~\cite{CO2} also shows that $Z[S^2\times S^2]\ne0$. Since $Z[S^3]$ is one-dimensional, gluing along $S^3$ yields $Z[X\mathbin{\#}Y]=Z[X]Z[Y]/Z[S^4]$. The stable-diffeomorphism argument of Ref.~\cite{CO2} then gives, for every simply connected spin four-manifold $X$,
\begin{equation}
    |Z[X]|^2=Z[S^2\times S^2]^{\chi(X)-2}Z[S^4]^{4-\chi(X)}\ne0.
\end{equation}
Since K3 is simply connected and spin, its partition function is nonzero. This contradicts the conclusion above, so a quantum theory with anomaly $\alpha$ cannot flow to a symmetry-preserving topological order of this kind. If $s\ne0$, the $p+ip$ class restricts nontrivially to the unitary subgroup $\widetilde G=\ker(s)$, and the same argument applied to $\widetilde G$ gives the result; see Ref.~\cite[Proposition 3.31]{Debray:2025kfg}.

\section{Applications}\label{sec:applications}
Our results constrain the dynamics of quantum theories in 3 spatial dimensions. As our first application, consider a gauge theory with $N_f$ flavors of Dirac fermions transforming in some representation $R$ of the gauge group $\mathcal{G}$, with the Lagrangian taking the standard form~\cite{peskin2018introduction}:
\begin{equation}\label{eq:Lagrangian}
\mathcal{L} = \sum_{i = 1}^{N_f} i \bar{\psi}_i \slashed{D}_R\psi_i,\quad \slashed{D}_R = \gamma^\mu(\partial_\mu + i A_\mu^a t_R^a),
\end{equation}
where $\psi_i$ are charged Dirac fermions, $A_\mu^a$ are $\mathcal{G}$ gauge fields, and $t_R^a$ are the generators of the Lie algebra of $\mathcal{G}$ in the representation $R$. If the fermion parity $(-1)^F$ is not part of $\mathcal{G}$, local fermions are present, and our discussion applies. Importantly, we allow symmetry-preserving interactions and ask whether the anomaly forbids a symmetry-preserving gap anywhere in the phase diagram. We also identify topological orders that provide candidate anomaly matches when such a gap is not forbidden. This is reminiscent of ``symmetric mass generation'' \cite{Wang:2022ucy,Razamat:2020kyf} when no 't Hooft anomaly is present, and is sometimes referred to as ``topological mass generation''. We derive these constraints solely from the selected global symmetries and their corresponding 't Hooft anomalies. They are direct consequences of Theorems~\ref{thm:always_triv} and~\ref{thm:gapless} and Table~\ref{tab:results_TQFT}.

Classically, there is a chiral $\U(1)^F$ symmetry rotating the fermions, but the ABJ anomaly reduces this symmetry to
\begin{equation}\label{eq:abj-discrete-symmetry}
    \U(1)^F \rightarrow \Z_{{4 N_f \cdot T(R)}}^F,
\end{equation}
where $T(R)$ is the Dynkin index of the representation. In the convention of Ref.~\cite[Eq.~(8)]{CO2}, $I(R)=2T(R)$, and each Dirac fermion counts as two Weyl fermions. Relevant 't Hooft anomaly and supercohomology classifications are summarized in Table~\ref{tab:classification} and Appendix~\ref{app:discrete-chiral-anomalies}. To determine the precise anomaly class, write each Dirac fermion as two left-handed Weyl fermions. Each has unit charge under the chiral symmetry in Eq.~\eqref{eq:abj-discrete-symmetry}, so the total number of unit-charge Weyl fermions is
\begin{equation}
    N_{\mathrm W}=2N_f\dim(R).
    \label{eq:number-unit-weyl}
\end{equation}
In the coordinates chosen in Appendix~\ref{app:discrete-chiral-anomalies}, the generator of the first cyclic factor is normalized so that one unit-charge Weyl fermion has anomaly
\begin{equation}
    1\in\Z_{16}\quad\text{for }\Z_4^F,\qquad
    (1,0)\in\Z_{32}\oplus\Z_2\quad\text{for }\Z_8^F,\qquad
    1\in\Z_{144}\quad\text{for }\Z_{12}^F.
    \label{eq:unit-weyl-discrete-anomalies}
\end{equation}
Then we obtain the entries displayed in Table~\ref{tab:results_gaugetheories}.

\begin{table*}
\centering
\renewcommand{\arraystretch}{1.3}
\resizebox{\textwidth}{!}{%
\begin{tabular}
{c | c | c| c|c | c}
\hline \hline 
Gauge group $\mc{G}$ & Representation $R$ & Symmetry & 't Hooft anomaly & Symmetric gapped realization & Candidate anomaly match\\ \hline\hline
$\SU(2N+1)$ & fundamental & $\Z^F_2$ & trivial & Yes & Trivial gapped state\\ \hline
$\SO(2N+1)$ & vector & $\Z^F_4$& $2(2N+1)\in \Z_{16}$ & Yes &  $\Z_4$ gauge theory\\ \hline
$\SU(2)$ & $\boldsymbol 3$ &  $\Z^F_{8}$& $(6, 0) \in \Z_{32}\oplus\Z_2$ & No & No such TQFT\\ \hline
$\mathrm{F}_4$ & fundamental & $\Z^F_{12}$& $52 \in \Z_{144}$ & No & No such TQFT\\
\hline  
\hline
\end{tabular}}
\caption{Examples of gauge theories with gauge groups $\mc{G}$ and ${N}_f=1$ Dirac fermion in representations $R$. The column titled ``Symmetry'' shows the global symmetries of these theories. The fourth column gives the group of possible 't Hooft anomalies for each symmetry together with the exact anomaly class of the theory. The last two columns indicate whether the anomaly admits a symmetry-preserving gapped realization and, when it does, give a candidate anomaly-matching realization from Table~\ref{tab:results_TQFT}.}
\label{tab:results_gaugetheories}
\end{table*}

We collect the results for the gauge groups $\mc{G}$ and representations $R$ considered here in Table~\ref{tab:results_gaugetheories}, including their selected global symmetries, associated 't Hooft anomalies, and anomaly-based IR constraints. In two examples, the 't Hooft anomalies lie beyond supercohomology. Within the class of IR descriptions specified above, the IR theory must consequently be gapless, or the chiral symmetry must be spontaneously broken, independent of any symmetry-preserving interactions one might add. In the other two examples, the anomaly does not forbid a symmetry-preserving gap, and candidate anomaly-matching gapped phases are listed in Table~\ref{tab:results_TQFT}. In particular, for the $\mathrm{SO}(2N+1)$ gauge theory with charged fermions in the vector representation, the anomaly cannot be matched by a trivial gapped phase or by the simplest $\Z_2$ gauge theory.

These results have direct consequences for physics beyond the Standard Model (BSM) \cite{2020arXiv200616996W,Cheng:2024awi,Wang:2025oow,Wan:2024kaf,Wang:2020mra}, where gauged chiral fermions play a significant role. It has been proposed in Ref.~\cite{Wang:2020mra} that the apparent incompleteness of the Standard Model—most notably the quantum anomalies arising from the absence of observed right-handed neutrinos—can be theoretically resolved by augmenting the SM with a nonperturbative topological sector. The specific topological orders investigated in this work provide concrete candidates for such a sector, thereby supporting the consistency of a fully gauged, anomaly-free formulation of the Standard Model without requiring conventional sterile neutrinos. These possibilities are analyzed further in more recent work by Wan, Wang, and Yau~\cite{Wang:2024auy,Wang:2025oow,Wan:2025ymd,ZJSNew}.

It is also interesting to analyze Weyl fermions with the Lagrangian in Eq.~\eqref{eq:Lagrangian_Weyl}. To introduce nontrivial dynamics, we can couple an extra bosonic field to the Weyl fermion $\chi$, and thus go beyond chiral symmetries. As a minimal example, consider a single left-handed Weyl fermion $\chi$ coupled to a bosonic field $\phi$, each carrying charge $1$ under a given $\Z_4$ symmetry. The combined system exhibits a $G_f = \Z_4\times \Z_2^F$ symmetry. The simplest nontrivial symmetric coupling that we can introduce is
\beq\label{eq:interaction}
\mathcal{L}_{\mathrm{int}} = g\phi^2\chi^T i\sigma_2\chi + \hc
\eeq
The anomaly $\alpha$ of this theory is an element of $\mho^5(\Z_4\times\Z_2^F)\cong\Z_4$~\cite[Example 7.3.3]{BG10}. Since the bosonic field does not contribute to the fermionic anomaly, the total system exhibits the same anomaly as the Weyl fermion, which carries anomaly $1\in \Z_4$~\cite[Eq.~(2.31)]{Hsi18}. Thus, the anomaly $\alpha$ is a beyond-supercohomology anomaly, as shown in Table~\ref{tab:classification}. From \cref{thm:gapless}, we conclude that neither the interaction in Eq.~\eqref{eq:interaction} nor any other symmetry-preserving interaction can produce a symmetric gapped phase in the TQFT class specified above without breaking $G_f$. It is worth noting that the $\Z_4\times \Z_2^F$ symmetry cannot be embedded into a chiral $\U(1)^F$ symmetry and therefore lies outside the class of discrete chiral symmetries discussed earlier in Sec.~\ref{sec:applications}. More generally, \cref{thm:gapless} implies that adding arbitrary bosonic degrees of freedom cannot produce a symmetric gapped phase in this class when the discrete chiral anomaly lies beyond supercohomology and the original fermionic symmetry is preserved.

In condensed matter systems, Weyl fermions emerge as low-energy quasiparticles near discrete band-crossing points in \emph{Weyl semimetals} \cite{2018RvMP...90a5001A,2025arXiv250401300Z}. Crucially, lattice symmetries may act nontrivially on these nodes, permuting the Weyl cones and generating nontrivial beyond-supercohomology anomalies. Our results constrain the phase diagram: if beyond-supercohomology anomalies are present, the system cannot enter a fully gapped phase in the TQFT class specified above without explicitly or spontaneously breaking the underlying lattice symmetries. Future numerical studies on lattice models would be valuable to corroborate these theoretical constraints.

\section{Summary}\label{sec:summary}
In this work, we established a systematic framework for constructing (3+1)d fermionic topological orders that saturate prescribed global anomalies. By generalizing the symmetry-extension procedure to finite symmetries in the fermionic setting, we identified a dichotomy: in-supercohomology anomalies can be saturated by gapped topological orders, while beyond-supercohomology anomalies cannot within the TQFT class specified above; hence, they imply symmetry-enforced gaplessness in that class. This result provides a rigorous criterion for constraining the symmetry-preserving IR possibilities of strongly coupled UV theories from their symmetry and anomaly data. 

Several future directions are of interest. First, we plan to extend our framework to continuous and time-reversal symmetries; in particular, for continuous symmetries, the criterion for symmetry-enforced gaplessness is expected to be notably more intricate and interesting \cite{2014PhRvB..89s5124W,Ning:2021kqc,Debray:2023bosonization}. Second, we will investigate the field-theoretic mechanisms by which parameter tuning drives the theory from candidate topological orders to chiral symmetry-breaking states \cite{PhysRev.122.345,Razamat:2020kyf}. Moreover, our results provide a pathway for constructing lattice models for chiral theories with anomalous discrete or continuous symmetries \cite{Gioia:2025bhl,Meyniel:2025euu,Thorngren:2026ydw}. We anticipate that the anomaly constrains how these symmetries act on the lattice in a specific non-on-site manner \cite{Gioia:2025bhl}. Finally, we wish to examine the relevance of our results to proposals for beyond-the-Standard-Model physics \cite{Wang:2020mra} in greater detail.

\section{Acknowledgements}\label{sec:acknowledgements}
It is a pleasure to thank
Clay Córdova,
Thibault Décoppet,
Jaume Gomis,
Weizhen Jia,
Theo Johnson-Freyd,
Ryohei Kobayashi,
Cameron Krulewski,
Tian Lan,
Miguel Montero,
Lukas Müller,
Kantaro Ohmori,
Luuk Stehouwer,
Zheyan Wan,
Chong Wang,
Juven Wang,
Rui Wen,
and Fei Zhou
for helpful conversations. We thank Zheyan Wan and Juven Wang for sharing their independent work \cite{Wan:2025ymd}, and Zheyan Wan, Juven Wang, and Shing-Tung Yau for sharing the follow-up work \cite{ZJSNew}. The overlap with their work is discussed in the introduction.

WY was supported by the Natural Sciences and Engineering Research Council of Canada (NSERC) and the European Commission under the Grant Foundations of
Quantum Computational Advantage. MY is supported by the EPSRC Open Fellowship EP/\allowbreak X01276X/1.

\appendix

\section{Discrete chiral anomalies}
\label{app:discrete-chiral-anomalies}

In this appendix, we summarize the calculation of discrete chiral anomalies for an arbitrary number of Weyl fermions in three spatial dimensions. For comprehensive analyses, see Refs.~\cite{Hsi18,Wan:2025lad}.

For a gauge-anomaly-free theory with $G\times \widetilde G$ symmetry,
gauging $\widetilde G$ does not change the pure $G$ anomaly provided
there is no mixed anomaly between $G$ and $\widetilde G$.  The gauge
theories in Sec.~\ref{sec:applications} are vectorlike and hence free of
gauge anomalies.  Their classical chiral $\U(1)^F$ symmetries have mixed
ABJ anomalies with the gauge symmetry; these mixed ABJ anomalies break each classical $\U(1)^F$ symmetry to
$\Z_{2m}^F$, where $m=2N_fT(R)$.  By construction, these surviving
subgroups have no remaining mixed gauge anomalies, so their pure 't Hooft
anomalies can be computed from the corresponding fermion spectra before
gauging.  After charge-conjugating the right-handed component, each
Dirac fermion gives two left-handed Weyl fermions of unit chiral charge.
Including the gauge-representation multiplicity, the total number of Weyl fermions is
\begin{equation}
N_{\mathrm W}=2N_f\dim(R).
\end{equation}

In the Dai--Freed framework \cite{Dai:1994kq}, the anomaly of a $G_f$-symmetric chiral
fermion theory in $n$ spatial dimensions is the obstruction to defining its partition function as
a symmetry-invariant complex number on every closed $(n+1)$-dimensional manifold with the variant of spin structure encoded by $G_f$ (see~\cite[\S 3.2]{Ste22}). This obstruction is encoded by the
exponentiated Atiyah--Patodi--Singer $\eta$-invariant. For $G_f = \Z_{2m}^F$ or $\Z_m\times\Z_2^F$, Refs.~\cite{Hsi18,Wan:2025lad} calculate these $\eta$-invariants on generating sets for the corresponding bordism groups, which suffices to determine the full anomaly.

Hsieh~\cite[\S 2.3]{Hsi18} constructs an isomorphism $\mho^5(\Z_{2m}^F)\cong\Z_{a_m}\oplus\Z_{b_m}$ for natural numbers $a_m,b_m$ depending on $m$.\footnote{In some cases, $a_m = 1$ or $b_m = 1$; in this case, $\Z_1$ means the trivial group.}
For the three symmetries used in Sec.~\ref{sec:applications},
these numbers are~\cite[(2.42)]{Hsi18}
\begin{equation}
    (a_2,b_2)=(16,1),\qquad
    (a_4,b_4)=(32,2),\qquad
    (a_6,b_6)=(144,1).
\end{equation}
To write the anomaly of a collection of Weyl fermions, choose odd
integer lifts $q_{L,R}$ of their $\Z_{2m}^F$ charges and define
\begin{equation}
\Delta q_1:=\sum_Lq_L-\sum_Rq_R,
\qquad
\Delta q_3:=\sum_Lq_L^3-\sum_Rq_R^3,
\end{equation}
where the minus signs account for the opposite chirality.  The charges
must be odd because the order-two element of $\Z_{2m}^F$ acts as
fermion parity.  The anomaly class below is independent of the chosen
integer lifts modulo $2m$ \cite{Hsi18,Wan:2025lad}.  
In the integer coordinates of Ref.~\cite{Wan:2025lad}, equivalent after
a change of cyclic generators to those of Ref.~\cite{Hsi18}, the
anomaly is the pair
\begin{equation}
(\widehat\nu_1,\widehat\nu_2)
\in
\Z_{a_m}\oplus\Z_{b_m},
\end{equation}
where 
\begin{equation}
\widehat\nu_1(m,\Delta q_3,\Delta q_1)
=
a_m
\frac{(2m^2+m+1)\Delta q_3-(m+3)\Delta q_1}{48m}
\pmod{a_m},
\end{equation}
and the formula for the second coordinate is 
\begin{equation}
    \widehat\nu_2(m,\Delta q_3,\Delta q_1)
= b_m\frac{(m + 1) \mathrm{gcd}(m + 1, 2)(\Delta q_3 - \Delta q_1)}{4m} \pmod{b_m}\,.
\end{equation}

It is convenient to normalize the cyclic generators so that one
left-handed Weyl fermion of charge $1$ has anomaly $(1,0)$.  For
$m=2$, the coordinates above already have this normalization.  For
$m=4$, the original coordinates of a charge-$1$ fermion are
\begin{equation}
\widehat\nu_{\Z_8^F}
=\bigl(\widehat\nu_1,\widehat\nu_2\bigr)
=(5,0)\in\Z_{32}\oplus\Z_2.
\end{equation}
Because $5^{-1}=13\pmod {32}$, normalized coordinates are
\begin{equation}
\nu_{\Z_8^F}
      :=\left(13\widehat\nu_1\pmod {32},\;
              \widehat\nu_2\pmod 2\right)
      \in \Z_{32}\oplus\Z_2.
\end{equation}
Similarly, for $m=6$, the original coordinate of a charge-$1$ fermion
is
\begin{equation}
\widehat\nu_{\Z_{12}^F}
=35\in\Z_{144}.
\end{equation}
Because $35^{-1}=107\pmod {144}$, the normalized coordinate is
\begin{equation}
\nu_{\Z_{12}^F}
      :=107\widehat\nu_{\Z_{12}^F}\pmod {144}.
\end{equation}
Substituting $\Delta q_3=\Delta q_1+24k$ into the general formulas and
applying these changes of coordinates gives
\begin{align}
    \nu_{\Z_4^F}&=\Delta q_1+12k\pmod {16},\notag\\
    \nu_{\Z_8^F}&=(\Delta q_1+4k\pmod {32},\,k\pmod 2),\notag\\
    \nu_{\Z_{12}^F}&=\Delta q_1+60k\pmod {144}.
    \label{eq:appendix-three-ak-formulas}
\end{align}
Here $k$ is an integer because $q^3-q\in24\Z$ for every odd integer $q$:
\begin{equation}
    k:=\frac{\Delta q_3-\Delta q_1}{24}\in\Z.
\end{equation}
In particular, $N$ left-handed Weyl fermions of charge $1$ have
$\Delta q_1=N$ and $k=0$, so their anomaly is obtained simply by
substituting $N$ in the first cyclic factor.

\bibliographystyle{JHEP}
\bibliography{reference}

@article{Cheng:2024awi,
  title = {Boundary topological orders of {($4+1$)d} fermionic {${\mathbb{Z}}_{2N}^{F}$} symmetry protected topological states},
  author = {Cheng, Meng and Wang, Juven and Yang, Xinping},
  journal = {Phys. Rev. B},
  volume = {113},
  issue = {12},
  pages = {125105},
  numpages = {13},
  year = {2026},
  month = {Mar},
  publisher = {American Physical Society},
  doi = {10.1103/bjvz-tmj2},
  url = {https://link.aps.org/doi/10.1103/bjvz-tmj2},
    eprint = "2411.05786",
    archivePrefix = "arXiv",
    primaryClass = "cond-mat.str-el",
}

@article{2020arXiv200616996W,
       author = {{Wang}, Juven},
        title = "{Anomaly and Cobordism Constraints Beyond the Standard Model: Topological Force}",
         year = 2020,
archivePrefix = {arXiv},
       eprint = {2006.16996},
 primaryClass = {hep-th},
       adsurl = {https://ui.adsabs.harvard.edu/abs/2020arXiv200616996W}
}

@article{CO1,
    author = "C\'ordova, Clay and Ohmori, Kantaro",
    title = "{Anomaly Obstructions to Symmetry Preserving Gapped Phases}",
    eprint = "1910.04962",
    archivePrefix = "arXiv",
    primaryClass = "hep-th",
    month = "10",
    year = "2019"
}

@article{CO2,
    author = "C\'ordova, Clay and Ohmori, Kantaro",
    title = "{Anomaly Constraints on Gapped Phases with Discrete Chiral Symmetry}",
    eprint = "1912.13069",
    archivePrefix = "arXiv",
    primaryClass = "hep-th",
    doi = "10.1103/PhysRevD.102.025011",
    journal = "Phys. Rev. D",
    volume = "102",
    number = "2",
    pages = "025011",
    year = "2020"
}

@article{Debray:2023ior,
    author = "Debray, Arun and Devalapurkar, Sanath K. and Krulewski, Cameron and Liu, Yu Leon and Pacheco-Tallaj, Natalia and Thorngren, Ryan",
    title = "{A long exact sequence in symmetry breaking: order parameter constraints, defect anomaly-matching, and higher Berry phases}",
    eprint = "2309.16749",
    archivePrefix = "arXiv",
    primaryClass = "hep-th",
    doi = "10.1007/JHEP07(2025)007",
    journal = "J. High Energ. Phys.",
    volume = "07",
    pages = "007",
    year = "2025"
}

@article{Debray:2023bosonization,
    author = "Debray, Arun and Ye, Weicheng and Yu, Matthew",
    title = "{Bosonization and Anomaly Indicators of (2+1)-D Fermionic Topological Orders}",
    eprint = "2312.13341",
    archivePrefix = "arXiv",
    primaryClass = "math-ph",
    doi = "10.1007/s00220-025-05344-z",
    journal = "Commun. Math. Phys.",
    volume = "406",
    number = "8",
    pages = "178",
    year = "2025"
}

@article{Dai:1994kq,
    author = "Dai, Xianzhe and Freed, Daniel S.",
    title = "{Eta invariants and determinant lines}",
    eprint = "hep-th/9405012",
    archivePrefix = "arXiv",
    doi = "10.1063/1.530747",
    journal = "J. Math. Phys.",
    volume = "35",
    pages = "5155--5194",
    year = "1994",
    note = "[Erratum: J.Math.Phys. 42, 2343--2344 (2001)]. \arxref{hep-th/9405012}"
}

@article{Ye:2022bkx,
    author = "Ye, Weicheng and Zou, Liujun",
    title = "{Anomaly of $(2+1)$-dimensional symmetry-enriched topological order from $(3+1)$-dimensional topological quantum field theory}",
    eprint = {2210.02444},
    archivePrefix = "arXiv",
    primaryClass = "cond-mat.str-el",
    doi = "10.21468/SciPostPhys.15.1.004",
    journal = "SciPost Phys.",
    volume = "15",
    number = "1",
    pages = "004",
    year = "2023"
}

@article{Debray:2025iqs,
        AUTHOR = {Debray, Arun and Ye, Weicheng and Yu, Matthew},
     TITLE = {Global {S}tructure in the {P}resence of a {T}opological
              {D}efect},
   JOURNAL = {Comm. Math. Phys.},
  FJOURNAL = {Communications in Mathematical Physics},
    VOLUME = {407},
      YEAR = {2026},
    NUMBER = {8},
     PAGES = {Paper No. 159},
      ISSN = {0010-3616,1432-0916},
   MRCLASS = {81T45 (57K16 82B10)},
  MRNUMBER = {5099898},
       DOI = {10.1007/s00220-026-05629-x},
       URL = {https://doi.org/10.1007/s00220-026-05629-x},
archivePrefix = "arXiv",
       eprint = "2501.18399",
 primaryClass = "math-ph",
}

@Inbook{Hooft1980,
author="{'t Hooft}, G.",
editor="{'t Hooft}, G.
and Itzykson, C.
and Jaffe, A.
and Lehmann, H.
and Mitter, P. K.
and Singer, I. M.
and Stora, R.",
title="Naturalness, Chiral Symmetry, and Spontaneous Chiral Symmetry Breaking",
bookTitle="Recent Developments in Gauge Theories",
year="1980",
publisher="Springer US",
address="Boston, MA",
pages="135--157",
isbn="978-1-4684-7571-5",
doi="10.1007/978-1-4684-7571-5_9",
url="https://doi.org/10.1007/978-1-4684-7571-5_9"
}

@article{Ning:2021kqc,
    author = "Ning, Shang-Qiang and Mao, Bin-Bin and Li, Zhengqiao and Wang, Chenjie",
    title = "{Anomaly indicators and bulk-boundary correspondences for three-dimensional interacting topological crystalline phases with mirror and continuous symmetries}",
    eprint = "2105.02682",
    archivePrefix = "arXiv",
    primaryClass = "cond-mat.str-el",
    doi = "10.1103/PhysRevB.104.075111",
    journal = "Phys. Rev. B",
    volume = "104",
    number = "7",
    pages = "075111",
    year = "2021"
}

@ARTICLE{2014PhRvB..89s5124W,
       author = {{Wang}, Chong and {Senthil}, T.},
        title = "{Interacting fermionic topological insulators/superconductors in three dimensions}",
      journal = {Phys. Rev. B},
         year = 2014,
        month = may,
       volume = {89},
       number = {19},
          eid = {195124},
        pages = {195124},
          doi = {10.1103/PhysRevB.89.195124},
archivePrefix = {arXiv},
       eprint = {1401.1142},
 primaryClass = {cond-mat.str-el},
       adsurl = {https://ui.adsabs.harvard.edu/abs/2014PhRvB..89s5124W}
}

@article{Wang:2017txt,
    author = "Wang, Chong and Nahum, Adam and Metlitski, Max A. and Xu, Cenke and Senthil, T.",
    title = "{Deconfined quantum critical points: symmetries and dualities}",
    eprint = "1703.02426",
    archivePrefix = "arXiv",
    primaryClass = "cond-mat.str-el",
    doi = "10.1103/PhysRevX.7.031051",
    journal = "Phys. Rev. X",
    volume = "7",
    number = "3",
    pages = "031051",
    year = "2017"
}

@article{D8,
  title={The {M}orita Theory of Fusion 2-Categories},
  author={D{\'e}coppet, Thibault D.},
  journal =       "Higher Structures",
  volume =        "7",
  number =        "1",
  pages =         "234-292",
  year =          "2023",
 doi = "10.21136/HS.2023.07",
archivePrefix = {arXiv},
       eprint = {2208.08722},
 primaryClass = {math.CT},
}

@article{2023arXiv230708147F,
       author = {{Freed}, Daniel S.},
        title = "{What is an anomaly?}",
archivePrefix = {arXiv},
        year = 2023,
       eprint = {2307.08147},
 primaryClass = {hep-th},
       adsurl = {https://ui.adsabs.harvard.edu/abs/2023arXiv230708147F}
}

@article{D9,
   AUTHOR = {D\'{e}coppet, Thibault D.},
     TITLE = {Drinfeld centers and {M}orita equivalence classes of fusion
              2-categories},
   JOURNAL = {Compos. Math.},
  FJOURNAL = {Compositio Mathematica},
    VOLUME = {161},
      YEAR = {2025},
    NUMBER = {2},
     PAGES = {305--340},
      ISSN = {0010-437X,1570-5846},
   MRCLASS = {16D90 (18M15 18M20 18N25)},
  MRNUMBER = {4920627},
MRREVIEWER = {Jian\ Min\ Chen},
       DOI = {10.1112/S0010437X24007553},
       URL = {https://doi.org/10.1112/S0010437X24007553},
          eprint = "2211.04917",
    archivePrefix = "arXiv",
    primaryClass = "math.QA",
       
}

@article{Decoppet:2023bay,
    author = "D\'ecoppet, Thibault D. and Yu, Matthew",
    title = "{Fiber 2-Functors and Tambara\textendash{}Yamagami Fusion 2-Categories}",
    eprint = "2306.08117",
    archivePrefix = "arXiv",
    primaryClass = "math.CT",
    doi = "10.1007/s00220-025-05249-x",
    journal = "Commun. Math. Phys.",
    volume = "406",
    number = "3",
    pages = "64",
    year = "2025"
}

@article{2025arXiv251225069N,
       author = {{Ning}, Shang-Qiang and {Ren}, Xing-Yu and {Wang}, Qing-Rui and {Qi}, Yang and {Gu}, Zheng-Cheng},
        title = "{Classification of Interacting Topological Crystalline Superconductors in Three Dimensions and Beyond}",
         year = 2025,
        month = dec,
archivePrefix = {arXiv},
       eprint = {2512.25069},
 primaryClass = {cond-mat.str-el},
       adsurl = {https://ui.adsabs.harvard.edu/abs/2025arXiv251225069N}
}

@article{D11,
  title={Extension Theory and Fermionic Strongly Fusion 2-Categories},
  author={D{\'e}coppet, Thibault Didier},
  journal={SIGMA. Symmetry, Integrability and Geometry: Methods and Applications},
  volume={20},
  pages={092},
  year={2024},
  publisher={SIGMA. Symmetry, Integrability and Geometry: Methods and Applications},
doi = {10.3842/SIGMA.2024.092},
archivePrefix = {arXiv},
       eprint = {2403.03211},
 primaryClass = {math.CT},
  note = {with an Appendix by Thibault Didier Décoppet and Theo Johnson-Freyd}
}

@article{Decoppet:2024htz,
    author = "Décoppet, Thibault D. and Huston, Peter and Johnson-Freyd, Theo and Nikshych, Dmitri and Penneys, David and Plavnik, Julia and Reutter, David and Yu, Matthew",
    title = "{The Classification of Fusion 2-Categories}",
    eprint = "2411.05907",
    archivePrefix = "arXiv",
    primaryClass = "math.CT",
    month = "11",
    year = "2024"
}

@book{peskin2018introduction,
  title={An Introduction to Quantum Field Theory},
  author={Peskin, Michael E. and Schroeder, Daniel V.},
  year={2018},
  publisher={CRC Press}
}

@article{PhysRevX.10.031055,
  title = {Construction and Classification of Symmetry-Protected Topological Phases in Interacting Fermion Systems},
  author = {Wang, Qing-Rui and Gu, Zheng-Cheng},
  journal = {Phys. Rev. X},
  volume = {10},
  issue = {3},
  pages = {031055},
  numpages = {64},
  year = {2020},
  month = {Sep},
  publisher = {American Physical Society},
  doi = {10.1103/PhysRevX.10.031055},
  url = {https://link.aps.org/doi/10.1103/PhysRevX.10.031055},
archivePrefix = {arXiv},
       eprint = {1811.00536},
 primaryClass = {cond-mat.str-el},
}

@book{Fre19,
    AUTHOR = {Freed, Daniel S.},
     TITLE = {Lectures on field theory and topology},
    SERIES = {CBMS Regional Conference Series in Mathematics},
    VOLUME = {133},
      NOTE = {Published for the Conference Board of the Mathematical
              Sciences},
 PUBLISHER = {American Mathematical Society, Providence, RI},
      YEAR = {2019},
     PAGES = {xi+186},
      ISBN = {978-1-4704-5206-3},
   MRCLASS = {57R56 (55-02 81Q70 81T45 82B10)},
  MRNUMBER = {3969923},
MRREVIEWER = {Arnav Tripathy},
       DOI = {10.1090/cbms/133},
       URL = {https://doi.org/10.1090/cbms/133},
}

@article{FH,
    author = "Freed, Daniel S. and Hopkins, Michael J.",
    title = "{Reflection positivity and invertible topological phases}",
    eprint = "1604.06527",
    archivePrefix = "arXiv",
    primaryClass = "hep-th",
    doi = "10.2140/gt.2021.25.1165",
    journal = "Geom. Topol.",
    volume = "25",
    pages = "1165--1330",
    year = "2021"
}

@article{Hason:2020yqf,
    author = "Hason, Itamar and Komargodski, Zohar and Thorngren, Ryan",
    title = "{Anomaly Matching in the Symmetry Broken Phase: Domain Walls, CPT, and the Smith Isomorphism}",
    eprint = "1910.14039",
    archivePrefix = "arXiv",
    primaryClass = "hep-th",
    doi = "10.21468/SciPostPhys.8.4.062",
    journal = "SciPost Phys.",
    volume = "8",
    number = "4",
    pages = "062",
    year = "2020"
}

@ARTICLE{JF,
       author = {{Johnson-Freyd}, Theo},
        title = "{On the Classification of Topological Orders}",
      journal = {Commun. Math. Phys.},
         year = 2022,
        month = jul,
       volume = {393},
       number = {2},
        pages = {989-1033},
          doi = {10.1007/s00220-022-04380-3},
archivePrefix = {arXiv},
       eprint = {2003.06663},
 primaryClass = {math.CT},
       adsurl = {https://ui.adsabs.harvard.edu/abs/2022CMaPh.393..989J}
}

@inbook{JF2,
    AUTHOR = {Johnson-Freyd, Theo},
     TITLE = {{$(3+1)D$} topological orders with only a {$\mathbb{Z}_2$}-charged
              particle},
 BOOKTITLE = {Quantum symmetries: tensor categories, {TQFT}s, and vertex
              algebras},
    SERIES = {Contemp. Math.},
    VOLUME = {813},
     PAGES = {175--210},
 PUBLISHER = {Amer. Math. Soc., [Providence], RI},
      YEAR = {2025},
   MRCLASS = {18M20 (55S10 55T10 81T45)},
  MRNUMBER = {4886204},
       DOI = {10.1090/conm/813/16288},
       URL = {https://doi.org/10.1090/conm/813/16288},
archivePrefix = {arXiv},
       eprint = {2011.11165},
 primaryClass = {math.QA},
}

@article{Kobayashi:2019lep,
    author = "Kobayashi, Ryohei and Ohmori, Kantaro and Tachikawa, Yuji",
    title = "{On gapped boundaries for SPT phases beyond group cohomology}",
    eprint = "1905.05391",
    archivePrefix = "arXiv",
    primaryClass = "cond-mat.str-el",
    reportNumber = "IPMU-19-0064",
    doi = "10.1007/JHEP11(2019)131",
    journal = "J. High Energ. Phys.",
    volume = "11",
    pages = "131",
    year = "2019"
}

@article{Meyniel:2025euu,
  title = {Equivalence class of emergent single Weyl fermion lattice models in three dimensions: Gapless superconductors and superfluids versus chiral fermions},
  author = {Meyniel, Gabriel and Zhou, Fei},
  journal = {Phys. Rev. B},
  volume = {113},
  issue = {18},
  pages = {184502},
  numpages = {23},
  year = {2026},
  month = {May},
  publisher = {American Physical Society},
  doi = {10.1103/734v-fryg},
  url = {https://link.aps.org/doi/10.1103/734v-fryg},
    eprint = "2510.25959",
    archivePrefix = "arXiv",
    primaryClass = "hep-th",
}

@ARTICLE{Loo:2025,
       author = {{Loo}, Kevin and {Wang}, Qing-Rui},
        title = "{Systematic construction of interfaces and anomalous boundaries for fermionic symmetry-protected topological phases}",
      journal = {Phys. Rev. B},
         year = 2025,
        month = may,
       volume = {111},
       number = {20},
          eid = {205102},
        pages = {205102},
          doi = {10.1103/PhysRevB.111.205102},
archivePrefix = {arXiv},
       eprint = {2412.18528},
 primaryClass = {cond-mat.str-el},
       adsurl = {https://ui.adsabs.harvard.edu/abs/2025PhRvB.111t5102L}
}

@article{Lan_2019,
       author = {{Lan}, Tian and {Wen}, Xiao-Gang},
        title = "{Classification of 3 +1 D Bosonic Topological Orders (II): The Case When Some Pointlike Excitations Are Fermions}",
      journal = {Phys. Rev. X},
         year = 2019,
        month = apr,
       volume = {9},
       number = {2},
          eid = {021005},
        pages = {021005},
          doi = {10.1103/PhysRevX.9.021005},
archivePrefix = {arXiv},
       eprint = {1801.08530},
 primaryClass = {cond-mat.str-el},
       adsurl = {https://ui.adsabs.harvard.edu/abs/2019PhRvX...9b1005L}
}

@article{SchommerPries:2018,
    author = {Schommer-Pries, Christopher J.},
    title = {Tori Detect Invertibility of Topological Field Theories},
    journal = {Geom. Topol.},
    volume = {22},
    number = {5},
    pages = {2713--2756},
    year = {2018},
    doi = {10.2140/gt.2018.22.2713},
    eprint = {1511.01772},
    archivePrefix = {arXiv},
    primaryClass = {math.AT},
}

@article{Tachikawa:2017gyf,
    author = "Tachikawa, Yuji",
    title = "{On gauging finite subgroups}",
    eprint = "1712.09542",
    archivePrefix = "arXiv",
    primaryClass = "hep-th",
    reportNumber = "IPMU-17-0183",
    doi = "10.21468/SciPostPhys.8.1.015",
    journal = "SciPost Phys.",
    volume = "8",
    number = "1",
    pages = "015",
    year = "2020"
}

@article{Wang:2017moj,
    author = "Wang, Qing-Rui and Gu, Zheng-Cheng",
    title = "{Towards a Complete Classification of Symmetry-Protected Topological Phases for Interacting Fermions in Three Dimensions and a General Group Supercohomology Theory}",
    eprint = "1703.10937",
    archivePrefix = "arXiv",
    primaryClass = "cond-mat.str-el",
    doi = "10.1103/PhysRevX.8.011055",
    journal = "Phys. Rev. X",
    volume = "8",
    number = "1",
    pages = "011055",
    year = "2018"
}

@article{Wang:2017loc,
    author = "Wang, Juven and Wen, Xiao-Gang and Witten, Edward",
    title = "{Symmetric Gapped Interfaces of SPT and SET States: Systematic Constructions}",
    eprint = "1705.06728",
    archivePrefix = "arXiv",
    primaryClass = "cond-mat.str-el",
    doi = "10.1103/PhysRevX.8.031048",
    journal = "Phys. Rev. X",
    volume = "8",
    number = "3",
    pages = "031048",
    year = "2018"
}

@book{wen2004quantum,
  title={Quantum field theory of many-body systems: From the origin of sound to an origin of light and electrons},
  author={Wen, Xiao-Gang},
  year={2004},
  publisher={Oxford University Press}
}

@ARTICLE{2024PhRvB.110w5117R,
       author = {{Ren}, Xing-Yu and {Ning}, Shang-Qiang and {Qi}, Yang and {Wang}, Qing-Rui and {Gu}, Zheng-Cheng},
        title = "{Stacking group structure of fermionic symmetry-protected topological phases}",
      journal = {Phys. Rev. B},
         year = 2024,
        month = dec,
       volume = {110},
       number = {23},
          eid = {235117},
        pages = {235117},
          doi = {10.1103/PhysRevB.110.235117},
archivePrefix = {arXiv},
    eprint = {2310.19058},
 primaryClass = {cond-mat.str-el},
       adsurl = {https://ui.adsabs.harvard.edu/abs/2024PhRvB.110w5117R}
}

@article{Chen:2011pg,
    author = "Chen, Xie and Gu, Zheng-Cheng and Liu, Zheng-Xin and Wen, Xiao-Gang",
    title = "{Symmetry protected topological orders and the group cohomology of their symmetry group}",
    eprint = "1106.4772",
    archivePrefix = "arXiv",
    primaryClass = "cond-mat.str-el",
    doi = "10.1103/PhysRevB.87.155114",
    journal = "Phys. Rev. B",
    volume = "87",
    number = "15",
    pages = "155114",
    year = "2013"
}

@ARTICLE{2013PhRvD..88d5013W,
       author = {{Wen}, Xiao-Gang},
        title = "{Classifying gauge anomalies through symmetry-protected trivial orders and classifying gravitational anomalies through topological orders}",
      journal = {Phys. Rev. D},
         year = 2013,
        month = aug,
       volume = {88},
       number = {4},
          eid = {045013},
        pages = {045013},
          doi = {10.1103/PhysRevD.88.045013},
archivePrefix = {arXiv},
       eprint = {1303.1803},
 primaryClass = {hep-th},
       adsurl = {https://ui.adsabs.harvard.edu/abs/2013PhRvD..88d5013W}
}

@article{Yang:2023gvi,
    author = "Yang, Xinping and Cheng, Meng",
    title = "{Gapped boundary of (4+1)d beyond-cohomology bosonic SPT phase}",
    eprint = "2303.00719",
    archivePrefix = "arXiv",
    primaryClass = "cond-mat.str-el",
    doi = "10.1103/PhysRevB.110.045137",
    journal = "Phys. Rev. B",
    volume = "110",
    number = "4",
    pages = "045137",
    year = "2024"
}

@article{Brennan:2023kpo,
    author = "Brennan, T. Daniel",
    title = "{Anomaly enforced gaplessness and symmetry fractionalization for Spin$_{G}$ symmetries}",
    eprint = "2308.12999",
    archivePrefix = "arXiv",
    primaryClass = "hep-th",
    doi = "10.1007/JHEP02(2024)065",
    journal = "J. High Energ. Phys.",
    volume = "02",
    pages = "065",
    year = "2024"
}

@article{Brennan:2023ynm,
    author = "Brennan, T. Daniel and Sheckler, Aiden",
    title = "{Anomaly enforced gaplessness for background flux anomalies and symmetry fractionalization}",
    eprint = "2311.00093",
    archivePrefix = "arXiv",
    primaryClass = "hep-th",
    doi = "10.1007/JHEP05(2024)159",
    journal = "J. High Energ. Phys.",
    volume = "05",
    pages = "159",
    year = "2024"
}

@article{DLM24,
    author = {DeLazzer Meunier, Adrien},
    title = {Arbitrary classes in {$>2$}-degree cohomology of a finite group with arbitrary coefficients may be trivialized in a finite extension},
    eprint = {2601.04374},
    archivePrefix = {arXiv},
    primaryClass = {math.GR},
    year = {2026},
}

@article{Debray:2025kfg,
    author = "Debray, Arun and Ye, Weicheng and Yu, Matthew",
    title = "{How to Build Anomalous (3+1)d Topological Quantum Field Theories}",
    eprint = "2510.24834",
    archivePrefix = "arXiv",
    primaryClass = "math-ph",
    month = "10",
    year = "2025"
}

@article{Wang:2022ucy,
    author = "Wang, Juven and You, Yi-Zhuang",
    title = "{Symmetric Mass Generation}",
    eprint = "2204.14271",
    archivePrefix = "arXiv",
    primaryClass = "cond-mat.str-el",
    doi = "10.3390/sym14071475",
    journal = "Symmetry",
    volume = "14",
    number = "7",
    pages = "1475",
    year = "2022"
}

@article{Wang:2020mra,
    author = "Wang, Juven",
    title = "{Unified model beyond grand unification}",
    eprint = "2012.15860",
    archivePrefix = "arXiv",
    primaryClass = "hep-th",
    doi = "10.1103/PhysRevD.103.105024",
    journal = "Phys. Rev. D",
    volume = "103",
    number = "10",
    pages = "105024",
    year = "2021"
}

@ARTICLE{2021PhRvR...3a3056C,
       author = {{Chen}, Yu-An and {Ellison}, Tyler D. and {Tantivasadakarn}, Nathanan},
        title = "{Disentangling supercohomology symmetry-protected topological phases in three spatial dimensions}",
      journal = {Phys. Rev. Research},
         year = 2021,
        month = jan,
       volume = {3},
       number = {1},
          eid = {013056},
        pages = {013056},
          doi = {10.1103/PhysRevResearch.3.013056},
     eprint = {2008.05652}
}

@article{2025arXiv250401300Z,
author = {Zhong, Mengyuan and Vu, Nam Thanh Trung and Zhai, Wenhao and Soh, Jian Rui and Liu, Yuanda and Wu, Jing and Suwardi, Ady and Liu, Huajun and Chang, Guoqing and Loh, Kian Ping and Gao, Weibo and Qiu, Cheng-Wei and Yang, Joel K. W. and Dong, Zhaogang},
title = {Weyl Semimetals: From Principles, Materials to Applications},
journal = {Advanced Materials},
volume = {37},
number = {37},
pages = {2506236},
doi = {10.1002/adma.202506236},
url = {https://advanced.onlinelibrary.wiley.com/doi/abs/10.1002/adma.202506236},
year = {2025},
archivePrefix = {arXiv},
       eprint = {2504.01300},
 primaryClass = {cond-mat.mtrl-sci},
       adsurl = {https://ui.adsabs.harvard.edu/abs/2025arXiv250401300Z}
}

@incollection{BC18,
    AUTHOR = {Beaudry, Agnès and Campbell, Jonathan A.},
     TITLE = {A guide for computing stable homotopy groups},
 BOOKTITLE = {Topology and quantum theory in interaction},
    SERIES = {Contemp. Math.},
    VOLUME = {718},
     PAGES = {89--136},
 PUBLISHER = {Amer. Math. Soc., [Providence], RI},
      YEAR = {2018},
   MRCLASS = {55Q10 (55R40 55S10 55T15)},
  MRNUMBER = {3869642},
        DOI = {10.1090/conm/718/14476},
    eprint = "1801.07530",
    archivePrefix = "arXiv",
    primaryClass = "math.AT",
}

@article{Cam17,
        author = {Jonathan A. Campbell},
        title = {Homotopy Theoretic Classification of Symmetry Protected Phases},
    year = 2017,
    eprint = "1708.04264",
    archivePrefix = "arXiv",
    primaryClass = "math.AT",
}

@incollection{DG18,
    AUTHOR = {Debray, Arun and Gunningham, Sam},
     TITLE = {The {A}rf-{B}rown {TQFT} of pin{$^-$} surfaces},
 BOOKTITLE = {Topology and quantum theory in interaction},
    SERIES = {Contemp. Math.},
    VOLUME = {718},
     PAGES = {49--87},
 PUBLISHER = {Amer. Math. Soc., Providence, RI},
      YEAR = {2018},
   MRCLASS = {57R56 (57R15 81T25)},
  MRNUMBER = {3869641},
MRREVIEWER = {Carlos Segovia Gonz\'{a}lez},
       DOI = {10.1090/conm/718/14478},
    eprint = "1803.11183",
    archivePrefix = "arXiv",
    primaryClass = "math-ph",
}

@article{Benini:2018reh,
    author = "Benini, Francesco and C{\'o}rdova, Clay and Hsin, Po-Shen",
    title = "{On 2-Group Global Symmetries and their Anomalies}",
    eprint = "1803.09336",
    archivePrefix = "arXiv",
    primaryClass = "hep-th",
    reportNumber = "SISSA 10/2018/FISI, SISSA-10-2018-FISI",
    doi = "10.1007/JHEP03(2019)118",
    journal = "J. High Energ. Phys.",
    volume = "03",
    pages = "118",
    year = "2019"
}

@ARTICLE{2019PhR...827....1S,
       author = {{Senthil}, T. and {Son}, Dam Thanh and {Wang}, Chong and {Xu}, Cenke},
        title = "{Duality between (2 + 1) d quantum critical points}",
      journal = {Phys. Rept.},
         year = 2019,
        month = sep,
       volume = {827},
        pages = {1-48},
          doi = {10.1016/j.physrep.2019.09.001},
archivePrefix = {arXiv},
       eprint = {1810.05174},
 primaryClass = {cond-mat.str-el},
       adsurl = {https://ui.adsabs.harvard.edu/abs/2019PhR...827....1S}
}

@article{COSY20,
  title = {Decorated {${\mathbb{Z}}_{2}$} symmetry defects and their time-reversal anomalies},
  author = {C\'ordova, Clay and Ohmori, Kantaro and Shao, Shu-Heng and Yan, Fei},
  journal = {Phys. Rev. D},
  volume = {102},
  issue = {4},
  pages = {045019},
  numpages = {15},
  year = {2020},
  month = {Aug},
  publisher = {American Physical Society},
  doi = {10.1103/PhysRevD.102.045019},
      eprint = "1910.14046",
    archivePrefix = "arXiv",
    primaryClass = "hep-th"
}

@article{Gioia:2025bhl,
  title = {Exact Chiral Symmetries of {$3+1\mathrm{D}$} {H}amiltonian Lattice Fermions},
  author = {Gioia, Lei and Thorngren, Ryan},
  journal = {Phys. Rev. Lett.},
  volume = {136},
  issue = {6},
  pages = {061601},
  numpages = {8},
  year = {2026},
  month = {Feb},
  publisher = {American Physical Society},
  doi = {10.1103/s5q2-317m},
  url = {https://link.aps.org/doi/10.1103/s5q2-317m},
    eprint = "2503.07708",
    archivePrefix = "arXiv",
    primaryClass = "cond-mat.str-el",
}

@article{Thorngren:2026ydw,
    author = "Thorngren, Ryan and Preskill, John and Fidkowski, Lukasz",
    title = "{Chiral Lattice Gauge Theories from Symmetry Disentanglers}",
    eprint = "2601.04304",
    archivePrefix = "arXiv",
    primaryClass = "hep-th",
    month = "1",
    year = "2026"
}

@article{PhysRev.122.345,
  title = {Dynamical Model of Elementary Particles Based on an Analogy with Superconductivity. {I}},
  author = {Nambu, Y. and Jona-Lasinio, G.},
  journal = {Phys. Rev.},
  volume = {122},
  issue = {1},
  pages = {345--358},
  numpages = {0},
  year = {1961},
  month = {Apr},
  publisher = {American Physical Society},
  doi = {10.1103/PhysRev.122.345},
  url = {https://link.aps.org/doi/10.1103/PhysRev.122.345}
}

@unpublished{Kit13,
        title= {On the Classification of Short-Range Entangled States},
        author = {Alexei Kitaev},
        year = {2013},
        note= {Conference talk at the Simons Center. \url{http://scgp.stonybrook.edu/archives/7874}.}
}

@unpublished{Kit15,
        title= {Homotopy-theoretic approach to {SPT} phases in action: {$Z_{16}$} classification of three-dimensional superconductors},
        author = {Alexei Kitaev},
        year = {2015},
        note= {Conference talk at the Institute for Pure and Applied Mathematics. \url{http://www.ipam.ucla.edu/abstract/?tid=12389}}
}

@article{Kitaev:2005hzj,
    author = "Kitaev, Alexei",
    title = "{Anyons in an exactly solved model and beyond}",
    eprint = "cond-mat/0506438",
    archivePrefix = "arXiv",
    doi = "10.1016/j.aop.2005.10.005",
    journal = "Annals Phys.",
    volume = "321",
    number = "1",
    pages = "2--111",
    year = "2006"
}

@article{KT17,
       author = {{Kapustin}, Anton and {Thorngren}, Ryan},
        title = "{Fermionic SPT phases in higher dimensions and bosonization}",
      journal = {J. High Energ. Phys.},
         year = 2017,
        month = oct,
       volume = {2017},
       number = {10},
          eid = {80},
        pages = {80},
          doi = {10.1007/JHEP10(2017)080},
archivePrefix = {arXiv},
       eprint = {1701.08264},
 primaryClass = {cond-mat.str-el},
       adsurl = {https://ui.adsabs.harvard.edu/abs/2017JHEP...10..080K}
}

@article{Wan:2025lad,
    author = "Wan, Zheyan",
    title = "{Anomaly of 4D Weyl fermions with discrete symmetries}",
    eprint = "2506.19710",
    archivePrefix = "arXiv",
    primaryClass = "hep-th",
    doi = "10.1103/sq16-k99s",
    journal = "Phys. Rev. D",
    volume = "114",
    number = "3",
    pages = "036032",
    year = "2026"
}

@article{Wan:2024kaf,
    author = "Wan, Zheyan and Wang, Juven and You, Yi-Zhuang",
    title = "{Topological Responses of the Standard Model Gauge Group}",
    eprint = "2412.21196",
    archivePrefix = "arXiv",
    primaryClass = "hep-th",
    month = "12",
    year = "2024"
}

@article{Garcia-Etxebarria:2018ajm,
    author = "García Etxebarria, Iñaki and Montero, Miguel",
    title = "{Dai-Freed anomalies in particle physics}",
    eprint = "1808.00009",
    archivePrefix = "arXiv",
    primaryClass = "hep-th",
    reportNumber = "MPP-2018-188",
    doi = "10.1007/JHEP08(2019)003",
    journal = "J. High Energ. Phys.",
    volume = "08",
    pages = "003",
    year = "2019"
}

@article{Hsi18,
       author = {{Hsieh}, Chang-Tse},
        title = "{Discrete gauge anomalies revisited}",
         year = 2018,
archivePrefix = {arXiv},
       eprint = {1808.02881},
 primaryClass = {hep-th},
}

@article{ZJSNew,
        author = {Wan, Zheyan and Wang, Juven and Yau, Shing-Tung},
        title = {Fermion Families and {P}ontryagin Class: Topological Field Theory via Color Symmetry Extension},
        year = {2026},
        journal = {Phys. Rev. D},
        doi = {10.1103/jlf6-ccwl},
        archivePrefix = {arXiv},
        eprint = {2605.26202},
        primaryClass = {hep-th},
}

@article{2014arXiv1404.3230K,
       author = {{Kapustin}, Anton and {Thorngren}, Ryan},
        title = "{Anomalies of discrete symmetries in various dimensions and group cohomology}",
         year = 2014,
archivePrefix = {arXiv},
      eprint = {1404.3230},
 primaryClass = {hep-th},
}

@article{Thorngren:2015gtw,
    author = "Thorngren, Ryan and von Keyserlingk, Curt",
    title = "{Higher SPT's and a generalization of anomaly in-flow}",
    eprint = "1511.02929",
    archivePrefix = "arXiv",
    primaryClass = "cond-mat.str-el",
    month = "11",
    year = "2015"
}

@ARTICLE{2014PhRvL.112w1602K,
       author = {{Kapustin}, Anton and {Thorngren}, Ryan},
        title = "{Anomalous Discrete Symmetries in Three Dimensions and Group Cohomology}",
      journal = {Phys. Rev. Lett.},
         year = 2014,
        month = jun,
       volume = {112},
       number = {23},
          eid = {231602},
        pages = {231602},
          doi = {10.1103/PhysRevLett.112.231602},
archivePrefix = {arXiv},
       eprint = {1403.0617},
 primaryClass = {hep-th},
       adsurl = {https://ui.adsabs.harvard.edu/abs/2014PhRvL.112w1602K}
}

@article{Bhardwaj:2023kri,
    author = "Bhardwaj, Lakshya and Bottini, Lea E. and Fraser-Taliente, Ludovic and Gladden, Liam and Gould, Dewi S. W. and Platschorre, Arthur and Tillim, Hannah",
    title = "{Lectures on generalized symmetries}",
    eprint = "2307.07547",
    archivePrefix = "arXiv",
    primaryClass = "hep-th",
    doi = "10.1016/j.physrep.2023.11.002",
    journal = "Phys. Rept.",
    volume = "1051",
    pages = "1--87",
    year = "2024"
}

@article{Luo:2023ive,
    author = "Luo, Ran and Wang, Qing-Rui and Wang, Yi-Nan",
    title = "{Lecture notes on generalized symmetries and applications}",
    eprint = "2307.09215",
    archivePrefix = "arXiv",
    primaryClass = "hep-th",
    doi = "10.1016/j.physrep.2024.02.002",
    journal = "Phys. Rept.",
    volume = "1065",
    pages = "1--43",
    year = "2024"
}

@article{Cordova:2018acb,
    author = "C{\'o}rdova, Clay and Dumitrescu, Thomas T.",
    title = "{Candidate phases for SU(2) adjoint QCD$_4$ with two flavors from $\mathcal{N}=2$ supersymmetric Yang-Mills theory}",
    eprint = "1806.09592",
    archivePrefix = "arXiv",
    primaryClass = "hep-th",
    doi = "10.21468/SciPostPhys.16.5.139",
    journal = "SciPost Phys.",
    volume = "16",
    number = "5",
    pages = "139",
    year = "2024"
}

@book{BG10,
    AUTHOR = {Bruner, Robert R. and Greenlees, J. P. C.},
     TITLE = {Connective real {$K$}-theory of finite groups},
    SERIES = {Mathematical Surveys and Monographs},
    VOLUME = {169},
 PUBLISHER = {American Mathematical Society, Providence, RI},
      YEAR = {2010},
     PAGES = {vi+318},
      ISBN = {978-0-8218-5189-0},
   MRCLASS = {19L41 (13D45 19-02 19L47 19L64 55N15 55N91)},
  MRNUMBER = {2723113},
       DOI = {10.1090/surv/169},
       URL = {https://doi.org/10.1090/surv/169},
}

@article{Gia73a,
    AUTHOR = {Giambalvo, V.},
     TITLE = {Cobordism of line bundles with a relation},
   JOURNAL = {Illinois J. Math.},
  FJOURNAL = {Illinois Journal of Mathematics},
    VOLUME = {17},
      YEAR = {1973},
     PAGES = {442--449},
      ISSN = {0019-2082},
   MRCLASS = {57D85},
  MRNUMBER = {0321121},
MRREVIEWER = {E. C. Turner},
       URL = {https://projecteuclid.org/euclid.ijm/1256051610},
}

@article{Ben88,
    AUTHOR = {Benson, Dave},
     TITLE = {Spin modules for symmetric groups},
   JOURNAL = {J. London Math. Soc. (2)},
  FJOURNAL = {Journal of the London Mathematical Society. Second Series},
    VOLUME = {38},
      YEAR = {1988},
    NUMBER = {2},
     PAGES = {250--262},
      ISSN = {0024-6107},
   MRCLASS = {20C30},
  MRNUMBER = {966297},
MRREVIEWER = {A. Kerber},
       DOI = {10.1112/jlms/s2-38.2.250},
       URL = {https://doi.org/10.1112/jlms/s2-38.2.250},
}

@article{Bhardwaj:2023fca,
    author = "Bhardwaj, Lakshya and Bottini, Lea E. and Pajer, Daniel and Schäfer-Nameki, Sakura",
    title = "{Categorical Landau Paradigm for Gapped Phases}",
    eprint = "2310.03786",
    archivePrefix = "arXiv",
    primaryClass = "cond-mat.str-el",
    doi = "10.1103/PhysRevLett.133.161601",
    journal = "Phys. Rev. Lett.",
    volume = "133",
    number = "16",
    pages = "161601",
    year = "2024"
}

@article{Schafer-Nameki:2023jdn,
    author = "Schäfer-Nameki, Sakura",
    title = "{ICTP lectures on (non-)invertible generalized symmetries}",
    eprint = "2305.18296",
    archivePrefix = "arXiv",
    primaryClass = "hep-th",
    doi = "10.1016/j.physrep.2024.01.007",
    journal = "Phys. Rept.",
    volume = "1063",
    pages = "1--55",
    year = "2024"
}

@article{Wang:2025oow,
    author = "Wang, Juven",
    title = "{Topological Quantum Dark Matter via Global Anomaly Cancellation}",
    eprint = "2502.21319",
    archivePrefix = "arXiv",
    primaryClass = "hep-th",
    month = "2",
    year = "2025"
}

@article{Wan:2025ymd,
    author = "Wan, Zheyan and Wang, Juven",
    title = "{Anomalous (3+1)d Fermionic Topological Quantum Field Theories via Symmetry Extension}",
    eprint = "2512.25038",
    archivePrefix = "arXiv",
    primaryClass = "hep-th",
    month = "12",
    year = "2025"
}

@article{Wang:2024auy,
  title = {Topological leptogenesis},
  author = {Wang, Juven},
  journal = {Phys. Rev. Res.},
  volume = {8},
  issue = {2},
  pages = {L022024},
  numpages = {10},
  year = {2026},
  month = {May},
  publisher = {American Physical Society},
  doi = {10.1103/r763-lq1b},
  url = {https://link.aps.org/doi/10.1103/r763-lq1b},
    eprint = "2501.00607",
    archivePrefix = "arXiv",
    primaryClass = "hep-ph",
}

@article{Johnson-Freyd:2021tbq,
    author = "Johnson-Freyd, Theo and Yu, Matthew",
    title = "{Topological Orders in (4+1)-Dimensions}",
    eprint = "2104.04534",
    archivePrefix = "arXiv",
    primaryClass = "hep-th",
    doi = "10.21468/SciPostPhys.13.3.068",
    journal = "SciPost Phys.",
    volume = "13",
    number = "3",
    pages = "068",
    year = "2022"
}

@article{Yu:2020twi,
    author = "Yu, Matthew",
    title = "{Symmetries and anomalies of (1+1)d theories: 2-groups and symmetry fractionalization}",
    eprint = "2010.01136",
    archivePrefix = "arXiv",
    primaryClass = "hep-th",
    doi = "10.1007/JHEP08(2021)061",
    journal = "J. High Energ. Phys.",
    volume = "08",
    pages = "061",
    year = "2021"
}

@article{Gaiotto:2017zba,
    author = "Gaiotto, Davide and Johnson-Freyd, Theo",
    title = "{Symmetry Protected Topological phases and Generalized Cohomology}",
    eprint = "1712.07950",
    archivePrefix = "arXiv",
    primaryClass = "hep-th",
    doi = "10.1007/JHEP05(2019)007",
    journal = "J. High Energ. Phys.",
    volume = "05",
    pages = "007",
    year = "2019"
}

@article{Decoppet:2025eic,
    author = "D{\'e}coppet, Thibault D. and Yu, Matthew",
    title = "{The Classification of 3+1d Symmetry Enriched Topological Order}",
    eprint = "2509.10603",
    archivePrefix = "arXiv",
    primaryClass = "math-ph",
    month = "9",
    year = "2025"
}

@article{Stockall:2025ngz,
    author = "Stockall, Devon and Yu, Matthew",
    title = "{Geometric Categories for Continuous Gauging}",
    eprint = "2511.08254",
    archivePrefix = "arXiv",
    primaryClass = "math-ph",
    month = "11",
    year = "2025"
}

@article{Bhardwaj:2024qiv,
    author = "Bhardwaj, Lakshya and Pajer, Daniel and Schäfer-Nameki, Sakura and Tiwari, Apoorv and Warman, Alison and Wu, Jingxiang",
    title = "{Gapped phases in (2+1)d with non-invertible symmetries: Part I}",
    eprint = "2408.05266",
    archivePrefix = "arXiv",
    primaryClass = "hep-th",
    doi = "10.21468/SciPostPhys.19.2.056",
    journal = "SciPost Phys.",
    volume = "19",
    number = "2",
    pages = "056",
    year = "2025"
}

@ARTICLE{2018RvMP...90a5001A,
       author = {{Armitage}, N.~P. and {Mele}, E.~J. and {Vishwanath}, Ashvin},
        title = "{Weyl and Dirac semimetals in three-dimensional solids}",
      journal = {Rev. Mod. Phys.},
         year = 2018,
        month = jan,
       volume = {90},
       number = {1},
          eid = {015001},
        pages = {015001},
          doi = {10.1103/RevModPhys.90.015001},
archivePrefix = {arXiv},
       eprint = {1705.01111},
 primaryClass = {cond-mat.str-el},
       adsurl = {https://ui.adsabs.harvard.edu/abs/2018RvMP...90a5001A}
}

@article{Bhardwaj:2025piv,
    author = "Bhardwaj, Lakshya and Schäfer-Nameki, Sakura and Tiwari, Apoorv and Warman, Alison",
    title = "{Gapped Phases in (2+1)d with Non-Invertible Symmetries: Part II}",
    eprint = "2502.20440",
    archivePrefix = "arXiv",
    primaryClass = "hep-th",
    month = "2",
    year = "2025"
}

@article{Razamat:2020kyf,
    author = "Razamat, Shlomo S. and Tong, David",
    title = "{Gapped Chiral Fermions}",
    eprint = "2009.05037",
    archivePrefix = "arXiv",
    primaryClass = "hep-th",
    doi = "10.1103/PhysRevX.11.011063",
    journal = "Phys. Rev. X",
    volume = "11",
    number = "1",
    pages = "011063",
    year = "2021"
}

@article{GJOS17,
        author = "Gurski, Nick and Johnson, Niles and Osorno, Angélica and Stephan, Marc",
        doi = "10.2140/agt.2017.17.2763",
        fjournal = "Algebraic \& Geometric Topology",
        journal = "Algebr. Geom. Topol.",
        number = "5",
        pages = "2763--2806",
        publisher = "MSP",
        title = "Stable {P}ostnikov data of {P}icard {$2$}-categories",
    note = "\url{https://arxiv.org/abs/1606.07032}",
        volume = "17",
        year = "2017",
        eprint = "1606.07032",
    archivePrefix = "arXiv",
    primaryClass = "math.AT",
}

@article{Cr,
    author =        "Sjoerd E. Crans",
    title =         "Generalized Centers of Braided and Sylleptic Monoidal 2-Categories",
    journal =       "Advances in Mathematics",
    volume =        "136",
    year =          "1998",
    pages =         "183-223",
}

@article{BottSpiral,
    author = "Debray, Arun and Krulewski, Cameron and Stehouwer, Luuk",
    title = "{Unraveling the Bott spiral}",
    eprint = "2605.00316",
    archivePrefix = "arXiv",
    primaryClass = "math-ph",
    month = "5",
    year = "2026"
}

@article{Deb21,
	author = {Arun Debray},
	title = {Invertible phases for mixed spatial symmetries and the fermionic crystalline equivalence principle},
    eprint = "2102.02941",
    archivePrefix = "arXiv",
    primaryClass = "math-ph",
    month = "2",
    year = "2021"
}

@article{ABK21,
	author = {David Aasen and Parsa Bonderson and Christina Knapp},
	title = {Characterization and Classification of Fermionic Symmetry Enriched
  Topological Phases},
	year = {2021},
	note = {\url{https://arxiv.org/abs/2109.10911}},
     eprint = "2109.10911",
    archivePrefix = "arXiv",
    primaryClass = "cond-mat.str-el",
}

@article{BCHM22,
  title = {Classification of {$(2+1)$D} invertible fermionic topological phases with symmetry},
  author = {Barkeshli, Maissam and Chen, Yu-An and Hsin, Po-Shen and Manjunath, Naren},
  journal = {Phys. Rev. B},
  volume = {105},
  issue = {23},
  pages = {235143},
  numpages = {60},
  year = {2022},
  month = {Jun},
  publisher = {American Physical Society},
  doi = {10.1103/PhysRevB.105.235143},
note = {\url{https://arxiv.org/abs/2109.11039}},
 eprint = "2109.11039",
    archivePrefix = "arXiv",
    primaryClass = "cond-mat.str-el",
}

@article{FH20,
	     AUTHOR = {Freed, Daniel S. and Hopkins, Michael J.},
     TITLE = {Invertible phases of matter with spatial symmetry},
   JOURNAL = {Adv. Theor. Math. Phys.},
  FJOURNAL = {Advances in Theoretical and Mathematical Physics},
    VOLUME = {24},
      YEAR = {2020},
    NUMBER = {7},
     PAGES = {1773--1788},
      ISSN = {1095-0761},
   MRCLASS = {81T45 (55P10)},
  MRNUMBER = {4313235},
       DOI = {10.4310/ATMP.2020.v24.n7.a3},
       URL = {https://doi.org/10.4310/ATMP.2020.v24.n7.a3},
    note = {\url{https://arxiv.org/abs/1901.06419}},
 eprint = "1901.06419",
    archivePrefix = "arXiv",
    primaryClass = "math-ph",
}

@article{YY23,
    AUTHOR = {Yamashita, Mayuko and Yonekura, Kazuya},
     TITLE = {Differential models for the {A}nderson dual to bordism
              theories and invertible {QFT}'s, {I}},
   JOURNAL = {J. G\"{o}kova Geom. Topol. GGT},
  FJOURNAL = {Journal of G\"{o}kova Geometry Topology. GGT},
    VOLUME = {16},
      YEAR = {2023},
     PAGES = {1--64},
   MRCLASS = {55N22 (57R56 81T45)},
  MRNUMBER = {4641957},
  eprint = "2106.09270",
    archivePrefix = "arXiv",
    primaryClass = "math.AT",
    note = {\url{https://arxiv.org/abs/2106.09270}}
}

@article{Ste22,
	title = {Interacting {SPT} phases are not {Morita} invariant},
	volume = {112},
	issn = {0377-9017, 1573-0530},
	url = {https://arxiv.org/abs/2110.07408},
	doi = {10.1007/s11005-022-01559-6},
	language = {en},
	number = {3},
	urldate = {2023-04-05},
	journal = {Letters in Mathematical Physics},
	author = {Stehouwer, Luuk},
	month = jun,
	year = {2022},
	note = {\url{https://arxiv.org/abs/2110.07408}},
	pages = {64},
     eprint = "2110.07408",
    archivePrefix = "arXiv",
    primaryClass = "hep-th",
}

@article{Wan:2018djl,
    author = "Wan, Zheyan and Wang, Juven",
    title = "{Adjoint QCD$_4$, Deconfined Critical Phenomena, Symmetry-Enriched Topological Quantum Field Theory, and Higher Symmetry-Extension}",
    eprint = "1812.11955",
    archivePrefix = "arXiv",
    primaryClass = "hep-th",
    doi = "10.1103/PhysRevD.99.065013",
    journal = "Phys. Rev. D",
    volume = "99",
    number = "6",
    pages = "065013",
    year = "2019"
}

@article{GKSW,
    AUTHOR = {Gaiotto, Davide and Kapustin, Anton and Seiberg, Nathan and
              Willett, Brian},
     TITLE = {Generalized global symmetries},
   JOURNAL = {J. High Energy Phys.},
  FJOURNAL = {Journal of High Energy Physics},
      YEAR = {2015},
    NUMBER = {2},
     PAGES = {172},
      ISSN = {1126-6708,1029-8479},
   MRCLASS = {83E30},
  MRNUMBER = {3321281},
       DOI = {10.1007/JHEP02(2015)172},
       URL = {https://doi.org/10.1007/JHEP02(2015)172},
       note ={\url{https://arxiv.org/abs/1412.5148}}
}

\end{document}